\documentclass[aps,prc,twocolumn,floatfix,showpacs,preprintnumbers,amsmath,amssymb,nofootinbib,groupedaddress]{revtex4-2}
\usepackage{color}
\usepackage{graphicx}
\graphicspath{{figures/}} 
\usepackage{dcolumn}    
\usepackage{multirow}
\usepackage{bm}         
\usepackage{sidecap}
\usepackage{hyperref}   
\usepackage{amsmath}
\usepackage{graphicx}
\usepackage{float}
\usepackage{adjustbox}
\usepackage{caption}
\usepackage{subcaption}
\usepackage[T1]{fontenc}

\usepackage{mathtools} 
\usepackage[version=3]{mhchem} 

\usepackage[dvipsnames,usenames]{xcolor}
\usepackage{mathrsfs,natbib}
\usepackage{epsf,amssymb,amsbsy,amsfonts,amssymb,amsmath}
\usepackage{slashed}
\usepackage{comment}

\usepackage{hyperref}
\usepackage[justification=raggedright]{caption}
\definecolor{dark-red}{rgb}{0.,0.,0}
\definecolor{dark-blue}{rgb}{0.,0.,1}
\definecolor{medium-blue}{rgb}{0,0,1}
\hypersetup{
    colorlinks, linkcolor={dark-red},
    citecolor={dark-blue}, urlcolor={medium-blue}
}

\newcommand{\sat}{\mathrm{sat}}
\newcommand{\sym}{\mathrm{sym}}
\newcommand{\tov}{\mathrm{TOV}}
\newcommand{\MM}{\mathrm{MM}}
\newcommand{\NM}{\mathrm{NUC}}
\newcommand{\QM}{\mathrm{QM}}

\newcommand{\fopt}{\mathrm{PT}}
\newcommand{\tr}{\mathrm{tr}}

\begin{document}
%

\title{Impact of soft versus hard equations of state on hybrid stars}

\author{H. G\"{u}ven}
\affiliation{Universit\'e Paris-Saclay, CNRS/IN2P3, IJCLab, 91405 Orsay, France}
\affiliation{Physics Department, Yildiz Technical University, 34220 Esenler, Istanbul, Turkey}
\affiliation{Faculty of Engineering and Architecture, Istanbul Esenyurt University, 34510 Esenyurt, Istanbul, Türkiye}

\author{K. Bozkurt}
\affiliation{Physics Department, Yildiz Technical University, 34220 Esenler, Istanbul, Turkey}
\affiliation{Universit\'e Paris-Saclay, CNRS/IN2P3, IJCLab, 91405 Orsay, France}

\author{E. Khan}
\affiliation{Universit\'e Paris-Saclay, CNRS/IN2P3, IJCLab, 91405 Orsay, France}
\affiliation{Institut Universitaire de France (IUF)}

\author{J. Margueron}
\affiliation{International Research Laboratory on Nuclear Physics and Astrophysics, Michigan State University and CNRS, East Lansing, MI 48824, USA}

\date{\today}

%
\begin{abstract}

The nuclear stiffness as a dominant contribution of nuclear symmetry energy is analyzed on the properties of compact stars considering constraints from nuclear physics and astrophysics. 
A compact star can be a neutron star composed only of nuclear matter or a hybrid star with a quark core. Two typical models (soft and stiff) are considered for the nuclear equation of state, and for the hybrid one, a parameterized first-order phase transition approach, completed with a linear quark matter equation of state, is implemented. We show that the phase transition reduces the tension between GW170817 and NICER observations, 
and we illustrate the impact of nuclear stiffness for the understanding of the nature of the binary system in GW170817.
The implication on nuclear symmetry energy is also discussed. 
We also confirm our previous findings that the GW170817 waveform is best described as a binary HS with a low-density onset of stiff quark matter. This could also be interpreted as a quarkyonic cross-over.
\end{abstract}

\maketitle

\section{Introduction}

Neutron stars (NSs) are laboratories for the properties of supra-dense matter and large isospin asymmetries (excess of neutrons over protons), requiring expertise from nuclear physics, particle physics, and astrophysics~\cite{Annala:2020}. Terrestrial experiments are complementary since they probe the properties of supra-dense matter close to nuclear saturation densities $n_\sat$~\cite{Margueron:2018a} (the average density of atomic nuclei), or above $n_\sat$ but remaining close to isospin symmetry~\cite{Sorensen:2024}. Since it connects isospin symmetric matter from terrestrial experiments to neutron-rich matter in NSs, the symmetry energy is instrumental. In order to make progress on the understanding of supra-dense matter existing in nature, a theoretical approach, combining the complementary nuclear physics knowledge with the multi-messenger astrophysical observations~\cite{Ozel:2016}, e.g., LIGO/VIRGO’s observations of GW170817, GW190425 and GW190814~\cite{Abbott:2017, Abbott:2020_GW190425, Abbott:2020_GW190814}, and milli-second pulsars observations by NICER~\cite{Miller:2019, Riley:2019, Miller:2021, Riley:2021}, is necessary.

The gravitational wave event GW170817, also known as the "golden event", contributes to better constraining the dense matter equation of state (EoS)~\cite{Abbott:2017}. The tidal deformability~\cite{Hinderer:2008,Hinderer:2010,Damour:2009} inferred from this event is measured to be $70<\tilde{\Lambda} \lessapprox 500-720$~\cite{Abbott:2019,De:2018} at 90\% confidence interval, which requires the EoS to be soft (predicting low values for NS radii). On the contrary, radio astronomy has detected compact stars (CSs) with masses larger than $2M_\odot$, which implies that the high-density part of the EoS should be stiff enough to support the gravitational field generated by massive stars. Nuclear EoSs marginally satisfy these constraints~\cite{Guven:2020,Somasundaram:2021,Mondal:2023} and a possible solution is to introduce stiff quark matter (QM) within a first-order phase transition (FOPT)~\cite{Han:2019,Montana:2019}. In our study, we adopt the model of Zdunik et al.~\cite{Zdunik:2013}, which is a parameterized approach where the FOPT and the QM EoS are controlled by only three parameters, see also Refs.~\cite{Alford:2013,Chamel:2013}: the QM onset density $n_\fopt$, the latent heat at the phase transition $\Delta\epsilon_\fopt$, and the sound speed in QM $c_{s,\fopt}$. It gives access to thermodynamic quantities such as the chemical potential, the pressure, and the energy density as a function of the baryon density $n_b$. Note that a crossover transition such as the one induced by the quarkyonic model (QycM) can also be at the origin of a stiff EoS, see Refs.~\cite{McLerran:2019, Margueron2021} for instance. The QycM can, however, be masqueraded by the FOPT~\cite{Zdunik:2013,Alford:2013,Chamel:2013} with a low density $n_\fopt$~\cite{Somasundaram:2022b}. Therefore, we explore the FOPT approach, considering also low values for $n_\fopt$, close to and above the saturation density $n_\sat$ ($n_\sat\approx 0.155$~fm$^{-3}$, $\rho_\sat\approx 2.7\times10^{14}$~g/cm$^{3}$~\cite{Margueron:2018a}) 

In our study, we adopt two typical nucleon EoS: a soft one represented by the Skyrme SLy5 Hamiltonian~\cite{Chabanat:1998} and a stiff one represented by the relativistic PKDD Lagrangian~\cite{Long:2004}. The soft nucleon model is compatible with recent chiral effective field theory predictions for the symmetry energy~\cite{Drischler:2019}, with a value for the slope of the symmetry energy $L_{\sym,2}=48.3$~MeV. It is also compatible with a combined PREX-II and a set of properties of finite nuclei (binding energies, charge radii, and dipole polarizabilities) analysis from Ref.~\cite{Reinhard:2021}. The stiff nucleon model is compatible with the analysis of the PREX-II experiment from Ref.~\cite{Reed:2021} with $L_{\sym,2}=79.5$~MeV.

We implement the constraints from gravitational wave signals, radio astronomy, and nuclear physics within a Bayesian framework, similar to our previous analysis in Ref.~\cite{Guven:2023}, where a large number of models are considered, and a probability is assigned to each pair of model and constraint. We explore three cases associated with the binary systems in the GW170817 observation: i) two neutron stars (BNS), which implies that QM occurs at sufficiently large density for the binary system to remain on the nucleonic branch for the GW170817 observation, ii) a hybrid star (HS) together with a neutron star (HSNS), using the fact that the two systems in GW170817 can have different masses, the low mass system being associated to a NS and the high mass one to a HS, or iii) two-hybrid stars (BHS), which implies a low value for the QM onset density $n_\fopt$, or equivalently QycM~\cite{Somasundaram:2022b} as discussed before.

Our paper is organized as follows: In Sec.~\ref{TF}, the main theoretical models are introduced, namely the nucleonic metamodel~\cite{Margueron:2018b}, the first order phase transition and the linear QM EoS~\cite{Zdunik:2013}, the general relativistic TOV equations~\cite{Tolman:1939,Oppenheimer:1939} for spherical CSs generating masses and radii, and the associated pulsation equation~\cite{Hinderer:2008,Hinderer:2010,Damour:2009}, which provide the tidal deformabilities for a single CS, and then the effective tidal deformability for a binary system such as GW170817. In Sec.~\ref{sec:UQ} we present how uncertainties in the FOPT parameters are quantified and how constraints from nuclear physics and astrophysics are implemented. In this section, we also discuss the role of the nuclear symmetry energy on the properties of CSs. Finally, conclusions are presented in Sec.~\ref{Con}.

\section{Theoretical Framework} 
\label{TF}

We detail the modeling that reflects the properties of dense matter (crust EoS, nucleon meta-model, and FOPT to QM) as well as the general relativity equations to obtain CS's properties. In this way, EoSs are linked to global CS's properties such as masses, radii, and tidal deformabilities.

\subsection{Crust and outer core}
\label{NMM}

The Douchin-Haensel EoS~\cite{Douchin:2001} is considered for the crust EoS. This EoS is based on a compressible liquid-drop model employing SLy4 Skyrme nuclear interaction~\cite{Chabanat:1998}. A cubic spline is employed for the transition between the crust and the core, smoothly connecting the crust EoS below $0.1\,n_{\sat}$ to the core EoS above $n_{\sat}$. Further details about this connection are given in Ref.~\cite{Margueron:2018b}.

The outer core is described by a nucleon approach. Nucleon matter is described by the metamodel ELFc introduced in Ref.~\cite{Margueron:2018a}, similarly to our previous studies~\cite{Guven:2020,Guven:2023}, and we consider two typical nucleonic EoS: SLy5~\cite{Chabanat:1998} (soft) and PKDD~\cite{Long:2004} (stiff).

The internal energy per particle is defined as
\begin{equation}
\varepsilon_\MM(n_0,n_1)=t^{FG*}(n_0,n_1)+v(n_0,n_1).
\label{eq:mm}
\end{equation}
where the nucleon density is $n_0=n_n+n_p$ and $n_1=n_n-n_p$, $n_n$ and $n_p$ being the neutron and proton densities. The first term in Eq.~\eqref{eq:mm} is the in-medium kinetic energy density and the second term is the interaction potential. The in-medium kinetic energy is related to the non-relativistic Fermi gas (FG) as
\begin{eqnarray}
t^{FG*}(n_0,n_1)&=& \frac{t^{FG}_{\textrm{sat}}}{2}\bigg(\frac{n_0}{n_{\textrm{sat}}}\bigg)^{2/3} \bigg[\bigg(1+\kappa_{\textrm{sat}}\frac{n_0}{n_{\textrm{sat}}}\bigg)f_1(\delta) \nonumber\\
&&+\kappa_{\textrm{sym}}\frac{n_0}{n_{\textrm{sat}}}f_2(\delta) \bigg],
\label{eq:e8}
\end{eqnarray}
where $t^{FG}_{\textrm{sat}}=(3\hbar^2/10m_N)(3\pi^2/2)^{2/3}n^{2/3}_\textrm{sat}$ is the kinetic energy per nucleon in symmetric matter (SM) at saturation, where the nucleonic mass $m_N$ is taken identical for neutrons and protons $m_N=(m_n+m_p)/2=938.919$~MeV/c$^2$, giving $t^{FG}_{\textrm{sat}}\approx 22.1\enspace\textrm{MeV}$. The second term in Eq.~\eqref{eq:mm} is the interaction potential expressed as
\begin{equation}
v(n_0,n_1)= \sum_{a \geq 0}^N {\frac{1}{a!}(c_a^{\sat}+c_a^{\sym}\delta^2)x^au_a(x)},
\label{eq:e9}
\end{equation}
where the isospin asymmetry parameter is $\delta=n_1/n_0$, and the function $u_a(x)=1-(-3x)^{N+1-a}$exp$(-bn_0/n_{\textrm{sat}})$ imposes the zero density limit. We fix $b=10$ln$2\approx6.93$ as in Ref.~\cite{Margueron:2018b}. 

The functions $f_1$ and $f_2$ in Eq.~\eqref{eq:e8} are defined as~\cite{Margueron:2018a},
\begin{eqnarray}
f_1(\delta) &=& (1+\delta)^{5/3}+(1-\delta)^{5/3}, \label{eq:e10} \\
f_2(\delta) &=& \delta(1+\delta)^{5/3}-\delta(1-\delta)^{5/3}, \label{eq:e11}
\end{eqnarray}
and they represent the correction due to the Landau in-medium effective mass in symmetric and asymmetric matter (AM), see Ref.~\cite{Margueron:2018b} for more details. The parameters $\kappa_{\sat/\sym}$ in Eq.~\eqref{eq:e8} represent the properties of the in-medium Landau mass at saturation density in SM and in neutron matter (NM), since
\begin{eqnarray}
\kappa_{\sat}&=&\frac{m_N}{m_{\sat}^*}-1=\kappa_s,\thinspace\textrm{in SM}\, , \nonumber\\
\kappa_{\sym} &=& \frac{1}{2} \Bigg[\frac{m_N}{m^*_n}-\frac{m_N}{m^*_p}\Bigg]_{n_\sat}=\kappa_s-\kappa_v,\thinspace\textrm{in NM}\, . 
\label{eq:e12}
\end{eqnarray}
The parameter $\kappa_v$ is the enhancement factor of the Thomas-Reiche-Kuhn sum rule~\cite{Bohigas:1979}.
Fixing $\kappa_{\sat/\sym}$, the coefficients $c_a^{\sat/\sym}$ for the potential energy in Eq.~\eqref{eq:e9} are directly related to the nuclear empirical parameters (NEP) through a one-to-one correspondence, see Ref.~\cite{Margueron:2018b} for more details.

\begin{figure}[t]
\centering
\includegraphics[width=0.5\textwidth]{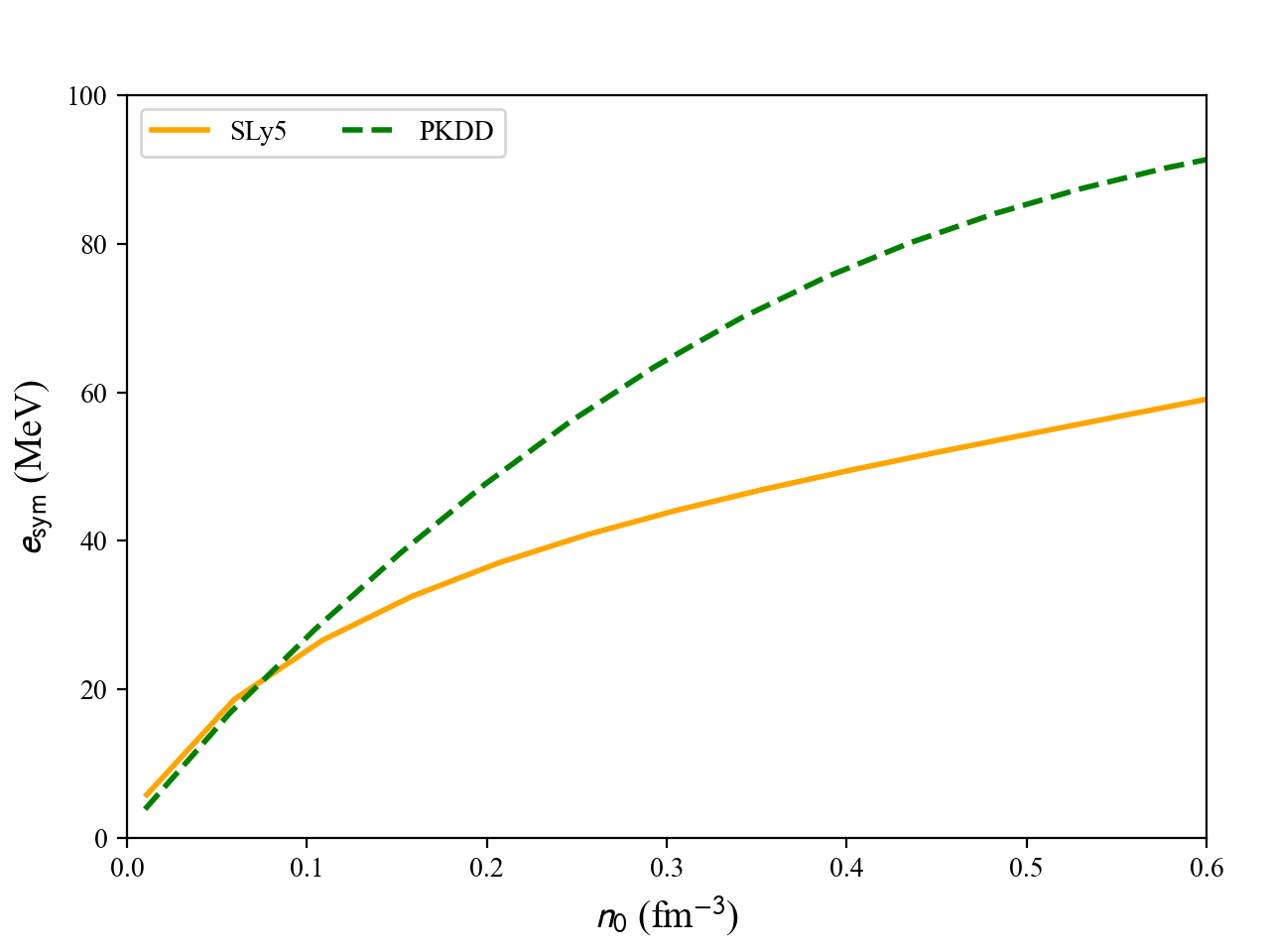}
\caption{Symmetry energy in MeV for SLy5~\cite{Chabanat:1998} and PKDD~\cite{Long:2004} as a function of the nucleon density $n_0$ in fm$^{-3}$.}
\label{fig:esym}
\end{figure}

The symmetry energy is defined as the difference between the neutron matter and symmetric matter energy per particle,
\begin{equation}
e_\sym(n_0) \equiv \frac{\epsilon_\MM(n_0,n_1=n_0)-\epsilon_\MM(n_0,n_1=0)}{n_0} \, .
\end{equation}
where $\varepsilon_{\MM}$ is the energy density for nucleons provided by the metamodel, see Eq.~\eqref{eq:mm}. The quadratic contribution to the symmetry energy is defined as
\begin{equation}
e_{\sym,2}(n_0) \equiv \frac {n_0}{2} \frac{\partial^2 \epsilon_\MM(n_0,n_1)}{\partial n_1^2}\vert_{n_1=0}\, ,
\end{equation}
See Ref.~\cite{Somasundaram:2021} for more details.

The density dependence of the nuclear symmetry energy $e_{\sym}$ is
shown in Fig.~\ref{fig:esym} for SLy5 (soft EoS) and PKDD (stiff EoS). 
Note that below saturation density, the difference between these two models is quite small. Above saturation density, however, the difference between the symmetry energy represent the present uncertainty band. 

\begin{table}[t]
\tabcolsep=0.2cm
\def\arraystretch{1.5}
\begin{tabular}{cccccc}
\hline\hline
EoS & $n_\sat$ & $E_\sym$ & $E_{\sym,2}$ & $L_\sym$ & $L_{\sym,2}$ \\
    & fm$^{-3}$ & MeV & MeV & MeV & MeV \\
\hline
SLy5 & 0.1604 & 32.72 & 32.03 & 49.8 & 48.3 \\
PKDD & 0.1500 & 37.94 & 36.79 & 93.4 & 79.5 \\
\hline\hline
\end{tabular}
\caption{Symmetry energy and slope of the symmetry energy for the nucleon models: SLy5~\cite{Chabanat:1998} and PKDD~\cite{Long:2004}. The saturation density $n_\sat$ and the quadratic contributions for the symmetry energy, $E_{\sym,2}$ and $L_{\sym,2}$, are obtained from the original papers~\cite{Chabanat:1998,Long:2004}, while the symmetry energy terms, $E_\sym$ and $L_\sym$, are obtained from \texttt{nucleardatapy} toolkit~\cite{Margueron:2025}.}
\label{tab:esym}
\end{table}

The soft nuclear EoS represented by Skyrme SLy5 Hamiltonian~\cite{Chabanat:1998} is compatible with recent chiral effective field theory predictions for the symmetry energy~\cite{Drischler:2019,Somasundaram:2021}, as well as binding energies, charge radii, and dipole polarisabilities for a set of finite nuclei~\cite{Reinhard:2021}. However, the tension with the PREX-II experiment on $r_\textrm{skin}^\textrm{Pb}$ still stands~\cite{Reinhard:2021}. The stiff nuclear EoS, PKDD~\cite{Long:2004}, is compatible with the analysis of the PREX-II experiment~\cite{Reed:2021}. See Table~\ref{tab:esym}, where the values of the symmetry energy parameters at saturation density $E_\sym$ and $L_\sym$, as well as their dominant quadratic contributions $E_{\sym,2}$ and $L_{\sym,2}$, are given.

Finally, the nuclear matter energy density $\epsilon_\NM$ includes the rest-mass term as,
\begin{equation}
\epsilon_\NM(n_0,n_1) = m_N c^2 \, n_0+\varepsilon_{\MM}(n_0,n_1) \, ,
\end{equation}
and the nuclear matter chemical potential is defined as
\begin{equation}
\mu_\NM(n_0) \equiv \frac{\partial \varepsilon_{\MM}(n_0,n_1)}{\partial n_0}\vert_{\beta}\, \, ,
\label{eq:chempot}
\end{equation}
at $\beta$ equilibrium. In terms of the neutron and proton chemical potentials, we have $\mu_\NM=x_n \mu_n + x_p\mu_p$, where the neutron and proton fractions are $x_q=n_q/n_0$ ($q=n$, $p$). The nuclear matter chemical potential $\mu_\NM$ is the quantity matching with QM chemical potential, see Sec.~\ref{sec:QEoS}.

\subsection{First order phase transition and quark matter equation of state}
\label{sec:QEoS}

The EoS in QM is provided by an approximation providing an analytic representation of modern EoS, see Ref.~\cite{Zdunik:2013} and references therein. Inspired by the MIT bag model~\cite{Chodos:1974}, QM is represented by a linear EoS, i.e., the pressure $p$ is a linear function of the energy density $\epsilon$. The energy density in nuclear and quark matter reads~\cite{Zdunik:2013,Alford:2013,Chamel:2013},
\begin{equation} 
\label{eq:e1}
\varepsilon(p) =\begin{cases}\varepsilon_{\NM}(p) \hbox{, if } p < p_\fopt, \\
\varepsilon_{\QM}(p)=\varepsilon_{\NM}(p_\fopt)+\Delta\varepsilon_\fopt+(p-p_\fopt)/ \alpha_\fopt \, , \end{cases}
\end{equation} 
otherwise, where $\alpha_\fopt$ is related to the sound speed in QM $c_{s,\fopt}$ as $\alpha_\fopt=(c_{s,\fopt}/c)^{2}$ and $c$ is the speed of light in the vacuum. Since the phase transition is of first order and the density is an order parameter, it is discontinuous through the phase transition and jumps from the density at which the first QM drop appears, hereafter referred to as $n_\fopt$, and the maximum density of the phase coexistence region, $n_\mathrm{tr}$. The density step in the FOPT is defined as, 
\begin{equation}
\Delta n_\fopt=n_\tr-n_\fopt \, ,
\end{equation}
producing the latent heat through the phase transition,
\begin{equation}
\Delta \epsilon_\fopt=\epsilon_\QM(n_\tr)-\epsilon_\NM(n_\fopt) \, .
\end{equation}
In addition, the following equalities are imposed between the pressure and the chemical potential at the boundaries of the FOPT: $p_\NM(n_\fopt)=p_\QM(n_\tr)=p_\fopt$ and $\mu_\NM(n_\fopt)=\mu_\QM(n_\tr)=\mu_\fopt$. Note that $p_\fopt$ is a model parameter, while $\mu_\fopt$ depends on the nuclear symmetry energy $e_\sym$.

In QM, the linear EoS \eqref{eq:e1} implies that the sound speed is constant, so the linear EoS is also referred to as the constant sound speed (CSS) approach. Note, however, that the QM sound speed is not necessarily constant: in the approach we adopt, it is sufficient that the QM sound speed is approximately constant close to the phase transition, and more precisely for a density range going from $n_\tr$ up to the maximal density probed by NS, $n_\tov$, or in terms of pressure from $p_\fopt$ up to $p_\tov$. Typical values for $n_\fopt$ and $n_\tr$ are 2-3$n_\sat$. The precise value of $p_\tov$ depends on the considered EoS. Above the maximal pressure $p_\tov$, variations of the sound speed from the constant approximation should not impact the properties of CS in the stable branch, which is where CSs are observed, nor the dynamical collapse to black holes when the unstable branch is explored, see discussion from Ref.~\cite{Ujevic:2024} for instance.

In the present approach, the description of the FOPT and the quark phase requires only three parameters: the pressure $p_\fopt$ at the onset of the phase transition, the shift in energy density $\Delta\varepsilon_\fopt$, reflecting the latent heat, and finally, the sound speed $c_{s,\fopt}$ in QM, which controls the stiffness of the quark EoS above the FOPT. 

In QM, the energy density $\varepsilon_{\QM}$ can be re-expressed from Eq.~\eqref{eq:e1} as
\begin{equation}
\epsilon_\QM(p) = \epsilon_\QM^* + \frac{p}{\alpha_\fopt} \, ,
\label{eq:eps}
\end{equation}
with
\begin{equation}
\epsilon^*_\QM=\epsilon_{\NM,\fopt}+\Delta\varepsilon_\fopt-p_\fopt/\alpha_\fopt \, ,
\label{eq:epsqp}
\end{equation}
where $\varepsilon_{\NM,\fopt}=\varepsilon_{\NM}(p_\fopt)$. 

In the following, the QM thermodynamic functions are expressed in terms of the baryon density $n_b$, which is associated with the conservation of the baryon number, i.e., $n_b=(n_u+n_d+n_s)/3$ for u-d-s QM. Note that the following relations are identical to the ones provided in the original Refs.~\cite{Zdunik:2013,Alford:2013,Chamel:2013}.

Since the sound speed parameter $\alpha_\fopt\equiv dp_\QM/d\epsilon_\QM=(n_b/\mu_\QM)dn_b/d\mu_\QM$, where the QM chemical potential $\mu_\QM\equiv d\epsilon_\QM/dn_b$ and the baryon density $n_b\equiv dp_\QM/d\mu_\QM$, we obtain
\begin{equation}
\mu_{\QM}(n_b)= \mu^*_{\QM} \left( \frac{ n_b}{n^*_{\QM}} \right)^{\alpha_\fopt} \, ,
\label{eq:mu}
\end{equation}
where $\mu^*_\QM$ and $n^*_\QM$ are parameters for QM EoS.

Since $\epsilon_\QM=n_b\mu_\QM-p_\QM$ (Euler relation), the pressure in QM can be expressed as,
\begin{equation}
\label{eq:pre}
p_{\QM}(n_b) = \varepsilon^*_{\QM} \frac{\alpha_\fopt}{1+\alpha_\fopt}\left[ \left( \frac{n_b}{n^*_{\QM}} \right)^{1+\alpha_\fopt}-1 \right],
\end{equation}
and using Eq.~\eqref{eq:eps}, the energy density in QM reads,
\begin{equation}
\label{eq:epsn}
\varepsilon_{\QM}(n_b) = \varepsilon^*_{\QM}\left\{1+\frac{1}{1+\alpha_\fopt} \left[ \left(\frac{n_b}{n^*_{\QM}} \right)^{1+\alpha_\fopt} -1\right]\right\}\, .
\end{equation}

The QM parameters $\varepsilon^*_{\QM}$, $\mu^*_\QM$, and $n^*_{\QM}$ are interpreted as the energy density, chemical potential, and the baryon density at zero QM pressure: $p_{\QM}=0$. Therefore, we obtain from the Euler relation, the following relation between these QM parameters,
\begin{equation}
n^*_{\QM} = \varepsilon^*_{\QM}/\mu^*_{\QM} \, ,
\label{eq:mustar}
\end{equation}
showing that fixing two of them is enough to determine the QM EoS. To parameterize the FOPT, the parameter $\mu^*_{\QM}$ can be employed instead of the latent heat $\Delta\varepsilon_\fopt$ since, 
\begin{equation} 
\label{eq:e2}
\Delta\varepsilon_\fopt = \frac{p_\fopt}{ \alpha_\fopt }  \left[ 1 + \frac{1 + \alpha_\fopt }{
\left[\frac{ \mu_{\NM,\fopt}}{\mu^*_\QM}\right]^{ \frac{1+\alpha_\fopt}{\alpha_\fopt}} -1} \right] - \varepsilon_{\NM,\fopt} \, ,
\end{equation}
using Eqs.~\eqref{eq:epsqp} and \eqref{eq:epsstar}, see hereafter.
In Eq.~\eqref{eq:e2}, $\Delta\varepsilon_\fopt$ is expressed as a function of the FOPT parameters $p_\fopt$ and $\alpha_\fopt$, the QM parameter $\mu_\QM^*$, and the nuclear matter properties $\mu_{\NM,\fopt}$ and $\varepsilon_{\NM,\fopt}$, which depend on the symmetry energy $e_\sym$.

In the following, the FOPT model is described in terms of the following three parameters: $p_\fopt$, $\mu_\QM^*$, and $\alpha_\fopt$. These parameters have a clear interpretation. For a given nuclear EoS, the parameter $p_\fopt$ is related to the density onset $n_\fopt$, the parameter $\mu^*_\QM$ is related to the QM EoS (for two-flavor color-superconducting model, $\mu^*_\QM\in[950,1000]$~MeV, and for color-flavor-locked superconducting model $\mu^*_\QM\in[1080,1150]$~MeV~\cite{Agrawal:2010}), and finally, the parameter $\alpha_{\fopt}$ reflects the quark interaction: without interaction, one expects $\alpha_{\fopt}=1/3$ (conformal limit), and quark interaction is repulsive for $\alpha_{\fopt} > 1/3$. 

Let us now invert Eq.~\eqref{eq:pre} to express the baryon density $n_b$ as a function of the pressure $p$:
\begin{equation}
n_b(p) = n^*_\QM\left[ 1+\frac{1+\alpha_\fopt}{\alpha_\fopt}\frac{p}{\epsilon^*_\QM}\right]^{1/(1+\alpha_\fopt)} \, ,
\label{eq:nb}
\end{equation}
from where one can obtain the QM parameter,
\begin{equation}
n_\QM^* = \frac{n_\fopt + \Delta n_\fopt}{\left[ 1+\frac{1+\alpha_\fopt}{\alpha_\fopt}\frac{p_\fopt}{\epsilon^*_\QM}\right]^{1/(1+\alpha_\fopt)} } \, ,
\label{eq:nstar}
\end{equation}
where $n_\fopt$ is fixed by $p_\fopt$ for a given nuclear EoS, and $\Delta n_\fopt$ is defined from Eq.~\eqref{eq:e2}. Therefore, the QM parameter $n_\QM^*$ is impacted by the symmetry energy $e_\sym$. Injecting Eq.~\eqref{eq:nb} into Eq.~\eqref{eq:mu}, the QM chemical potential becomes:
\begin{equation}
\mu_{\QM}(p)=\mu^*_{\QM} \left[1+ \frac{1+\alpha_\fopt}{\alpha_\fopt}\frac{p}{\varepsilon^*_{\QM}} \right]^{\frac{\alpha_\fopt}{1+\alpha_\fopt}} \, ,
\label{eq:muqp}
\end{equation}
recovering the expression for the chemical potential given in Ref.~\cite{Zdunik:2013}. Since $\mu_\fopt\equiv\mu_\QM(p_\fopt)$, one obtains from Eq.~\eqref{eq:muqp}, the following expression for the QM parameter $\varepsilon^*_{\QM}$:
\begin{equation}
\varepsilon^*_{\QM} = p_\fopt\frac{ 1+\alpha_\fopt}{\alpha_\fopt} \left[ \left(\frac{\mu_{\fopt}}{\mu^*_{\QM}}\right)^\frac{1+\alpha_\fopt}{\alpha_\fopt} -1 \right]^{-1} \, ,
\label{eq:epsstar}
\end{equation}
where $\varepsilon^*_{\QM}$ is defined from the FOPT parameters $p_\fopt$, $\mu^*_\fopt$, $\alpha_\fopt$ and the nuclear matter property $\mu_{\NM,\fopt}\equiv \mu_\fopt$. We conclude that the three QM parameters $\epsilon^*_\QM$, $\mu^*_\QM$, and $n^*_\QM$ are entirely controlled by the FOPT parameters and the nuclear EoS, in particular, the nuclear symmetry energy. Therefore, the nuclear symmetry energy has an important role in the determination of the QM EoS.

\begin{figure}
\centering
\begin{subfigure}{0.5\textwidth}
\includegraphics[width=1\textwidth]{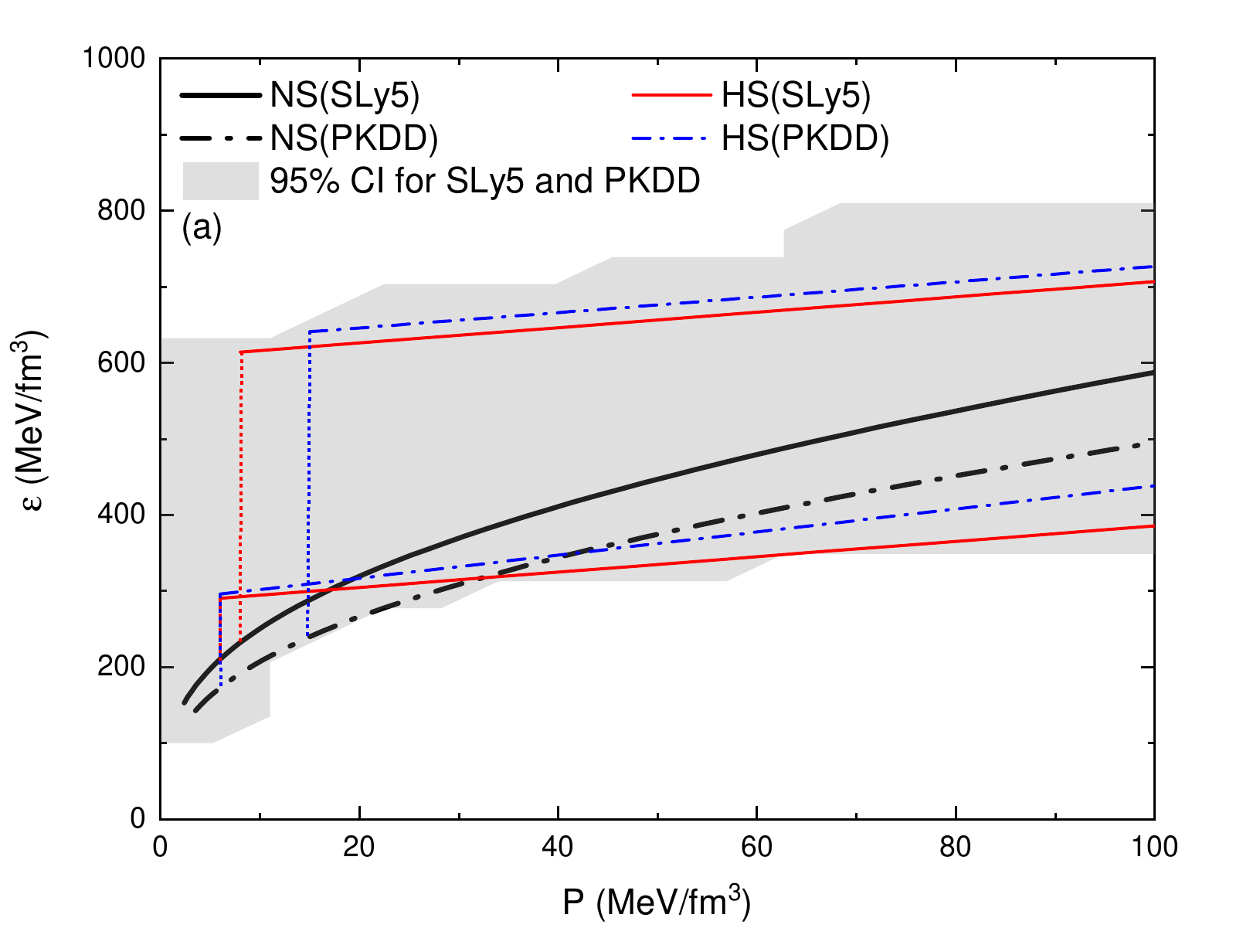}
\end{subfigure}
\begin{subfigure}{0.5\textwidth}
\includegraphics[width=1\textwidth]{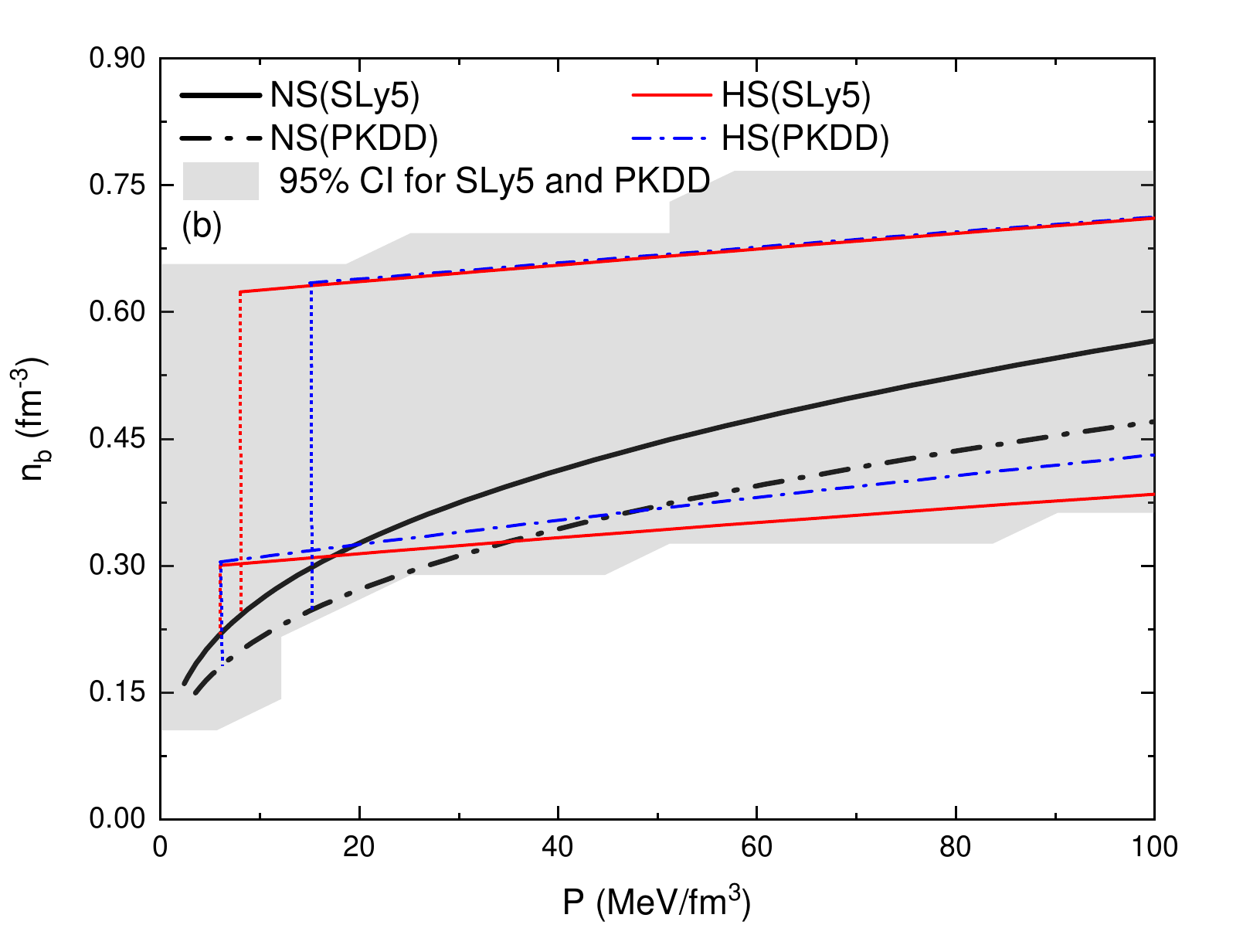}
\end{subfigure}
\begin{subfigure}{0.5\textwidth}
\includegraphics[width=1\textwidth]{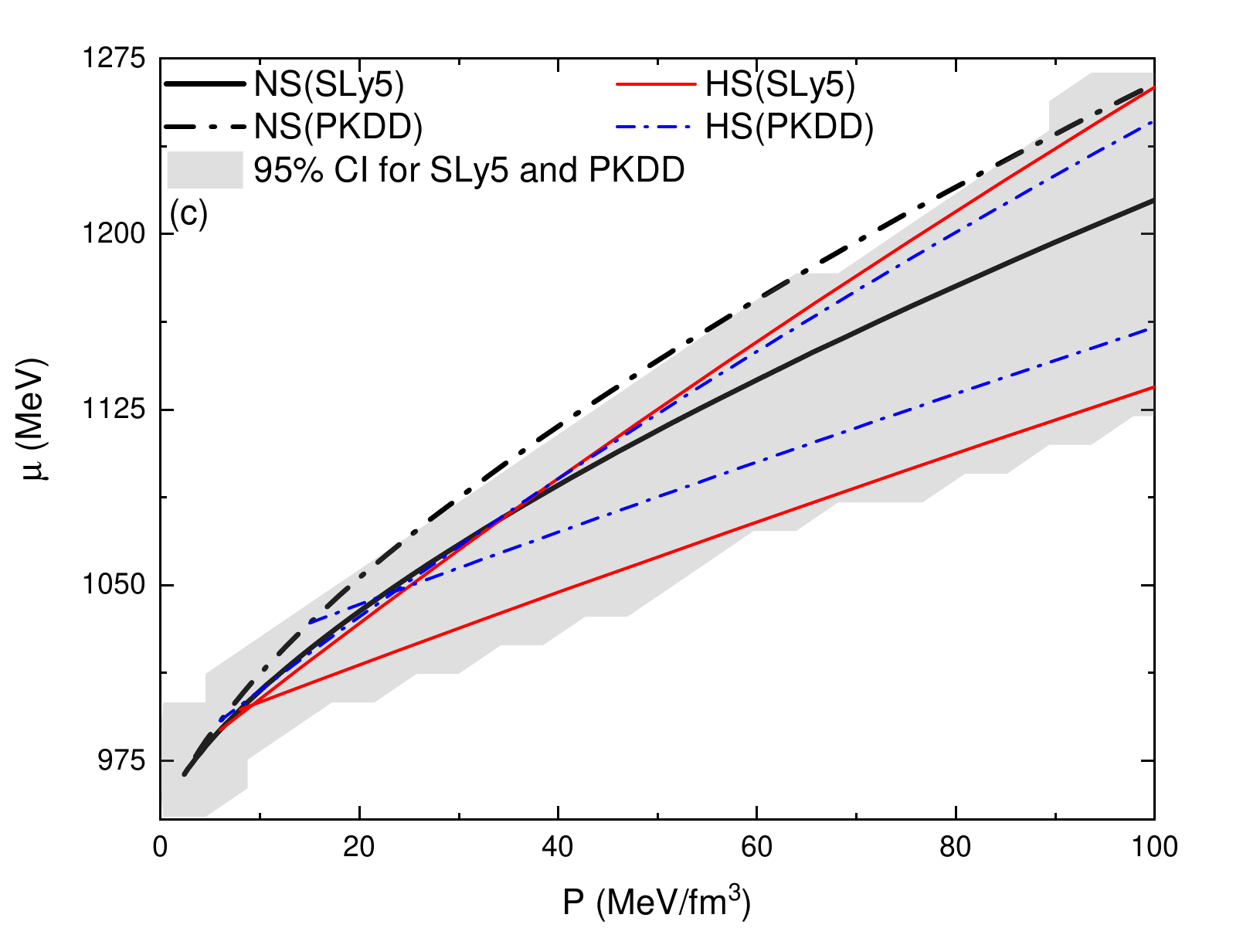}
\end{subfigure}
\caption{Energy-density (a), baryon density (b), and chemical potential (c) envelopes at 95\% credible interval for hybrid EoS based on PKDD and SLy5 nuclear model (grey area). The two nuclear EoS are shown in back lines while typical examples of hybrid EoS are shown in color lines (with a large and small density jump at the FOPT).}
\label{fig:EoSParameters}
\end{figure}

The energy density $\varepsilon$, the baryon density $n_\mathrm{b}$, and the chemical potential $\mu$ are shown in Fig.~\ref{fig:EoSParameters} (a), (b), and (c) as a function of the total pressure $p$. Note that in the nucleon phase, $n_\mathrm{b}=n_0+n_\mathrm{lep}$. A few examples of EoS are explicitly shown with curves in Fig.~\ref{fig:EoSParameters}. The nuclear EoSs are represented with back lines: the solid black line for SLy5 and the dot-dashed black line for PKDD. We show two examples of hybrid EoS for each nuclear model: one with a small density jump (strong FOPT) and another with a large density jump (weak FOPT). Note the continuity of the pressure and chemical potential shown in Fig.~\ref{fig:EoSParameters} (c), as expected for the FOPT. 

The grey areas in Fig.~\ref{fig:EoSParameters} represent the envelope at 95\% confidence interval for the exploration of all possible EoS, as described later in the uncertainty quantification of the model parameters~\ref{sec:UQ}.

\subsection{Neutron star global properties}
\label{sec:NSO}

The NS global properties such as masses, radii, and tidal deformabilities are deduced from the nuclear EoS assuming general relativity (TOV and pulsation equation)~\cite{Tolman:1939,Oppenheimer:1939,Hinderer:2008,Hinderer:2010,Damour:2009}. The solution of the hydrostatic equations in general relativity for spherical non-rotating and non-magnetic stars, also named the Tolman, Oppenheimer, Volkoff (TOV) equations are expressed as~\cite{Tolman:1939,Oppenheimer:1939}:
\begin{eqnarray}
&&\frac{dm(r)c^2}{dr} = 4\pi r^2\epsilon(r), \nonumber \\
&&\frac{dp(r)}{dr} = -\epsilon(r)\Bigg(1+\frac{p(r)}{\epsilon(r)} \Bigg)\frac{d\Phi(r)}{dr},  \label{eq:e14}\\
&&\frac{d\Phi(r)}{dr} = \frac{Gm(r)}{c^2 r^2}\Bigg(1+\frac{4\pi p(r) r^3}{m(r) c^2} \Bigg)\Bigg(1-\frac{2Gm(r)}{r c^2} \Bigg)^{-1}\hspace{-0.3cm}, 
\nonumber
\end{eqnarray}
where $G$ is the gravitational constant, $c$ the speed of light in vacuum, $p(r)$ the pressure and $m(r)$ the enclosed mass at radius r. Specifying the EoS in the crust and in the core, $p(\epsilon)$, the TOV system~\eqref{eq:e14} is closed, and a solution can exist. Note that $\epsilon(r)$ is the mass-energy density~\eqref{eq:e1} with additional contribution from leptons in the nuclear phase.

The wave-form signal obtained by LIGO-Virgo GW interferometers, see the analysis of GW170817 in Ref.~\cite{Abbott:2017} for instance, fixes several properties of the binary system, which are the masses of the merging CSs, $M_1$ and $M_2$ (by convention $M_1\ge M_2$ and the mass ratio $q=M_2/M_1\le 1$), as well as the effective tidal deformability, defined as
\begin{equation}
\label{e20}
\tilde{\Lambda} = \frac{16}{13}\frac{(M_1+12M_2)M_1^4\Lambda_1+(M_2+12M_1)M_2^4\Lambda_2}{(M_1+M_2)^5},
\end{equation}
where $\Lambda_1$ and $\Lambda_2$ are the tidal deformabilities of each of the CSs forming the binary system. In fact, the wave-form signal directly constrains the chirp mass, defined as 
\begin{equation}
M_\text{chirp} = (M_1M_2)^{3/5} M_\text{tot}^{1/5}\, ,
\end{equation}
where $M_\text{tot}=M_1+M_2$, and the effective tidal deformability is obtained at the fifth-order.

If $M_1=M_2$ in Eq.~\eqref{e20} then $\tilde{\Lambda}=\Lambda_1=\Lambda_2$. In the present study, we also consider the asymmetric case. 

The tidal deformability $\Lambda_i$ for a single NS can be expressed in terms of the Love number as~\cite{Hinderer:2008,Hinderer:2010}:
\begin{equation}
\label{e15}
\Lambda_i=\frac{2 k_{2,i}}{3 C_i^5},
\end{equation}
where the compactness of the CS is $C_i=(G M_i)/(c^2 R_i)$, for mass $M_i$ and radius $R_i$. The Love number $k_{2,i}$ is determined at the surface of the CS as,
\begin{eqnarray}
\nonumber  k_{2,i}&=& \frac{8C_i^5}{5}(1-2C_i)^2[2+2C_i(Y_i-1)-Y_i]  \\
\nonumber  &&\times  \big\{2C_i[6-3Y_i+3C_i(5Y_i-8)] \\
\nonumber  &&+4C_i^3[13-11Y_i+C_i(3Y_i-2)+2C_i^2(1+Y_i)]\\
\nonumber  &&+3(1-2C_i)^2[2-Y_i+2C_i(Y_i-1)] \\
&&\times \textrm{ln}(1-2C_i)  \big\} ^{-1},
\label{e16}
\end{eqnarray}
where $Y_i=y(R_i)$ is the solution of the pulsation equation for $r=R_i$~\cite{Postnikov:2010,Moustakidis:2017}:
\begin{equation}
\label{eq:e17}
r\frac{dy(r)}{dr}+y(r)^2+y(r)F(r)+Q(r)=0,
\end{equation}
with
\begin{eqnarray}
F(r) &=& \frac{1}{1-\frac{2Gm(r)}{rc^2}} \Bigg\{1+\frac{ 4\pi G r^2  }{c^4}\big(p(r)-\epsilon(r)\big)\Bigg\}, \label{e18}\\
Q(r)&=& \frac{4\pi G r^2/c^4}{1-\frac{2Gm(r)}{rc^2}}\Bigg\{5\epsilon(r)+9p(r)+\frac{p(r)+\epsilon(r)}{ (c_s/c)^2_\beta}\Bigg\} \nonumber \\
\nonumber &&\hspace{-1cm}-\frac{6}{1-\frac{2Gm(r)}{rc^2}} -\frac{4G^2}{r^2 c^8} \Bigg\{\frac{m(r)c^2+4\pi r^3 p(r)}{1-\frac{2Gm(r)}{rc^2}} \Bigg\}^2, \nonumber \\
\label{e19}
\end{eqnarray}
where $(c_s/c)^2_\beta=dp/d\epsilon$ is the sound speed in CS matter at beta-equilibrium. The pulsation equation~\eqref{eq:e17} is solved after the TOV equations~\eqref{eq:e14}, knowing the energy-density and the pressure radial profiles.

For hybrid stars (HS), the same TOV system~\eqref{eq:e14} and pulsation equation~\eqref{eq:e17} are solved to obtain the HS radius, mass, and tidal deformability. However, due to the sharp change in the energy-density at $p_\fopt$, see Eqs.~\eqref{eq:e1}, the numerical solution of the TOV equations~\eqref{eq:e14} is unstable and requires the introduction of a smoothing of the phase transition: the original energy density is replaced by a spline interpolation centered at $p_\fopt$, where the spline interpolation begins at $(p_\fopt - 5)$~MeV~fm$^{-3}$ and ends at $(p_\fopt + 5)$~MeV~fm$^{-3}$. The same procedure was applied in Refs.~\cite{Han:2019,Guven:2023} and it didn't change the mass-radius relation quantitatively: it was shown that changing the width of the spline in a small interval, chosen here to be fixed to 5~MeV~fm$^{-3}$, does not change CS observables~\cite{Han:2019}. In addition, since the numerical spline removes the discontinuity in the density or the pressure through the FOPT, the solution $y(r)$ for the pulsation equation~\eqref{eq:e17} does not need to be corrected at the FOPT, as suggested in Ref.~\cite{Postnikov:2010}.

\begin{figure}
\centering
\begin{subfigure}{0.5\textwidth}
\hspace{-0.40cm}
\includegraphics[width=1\textwidth]{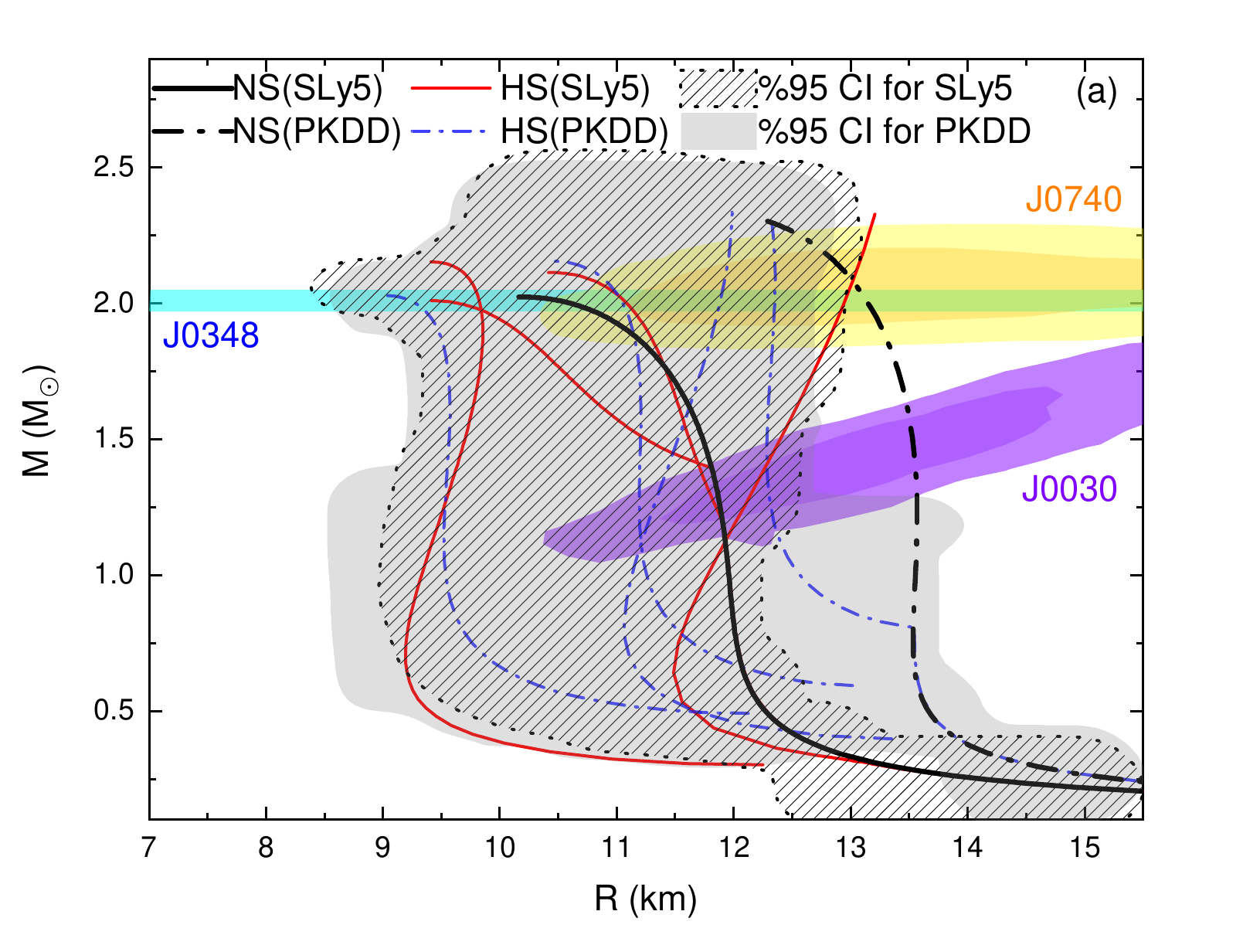}
\end{subfigure}
\begin{subfigure}{0.5\textwidth}
\includegraphics[width=1\textwidth]{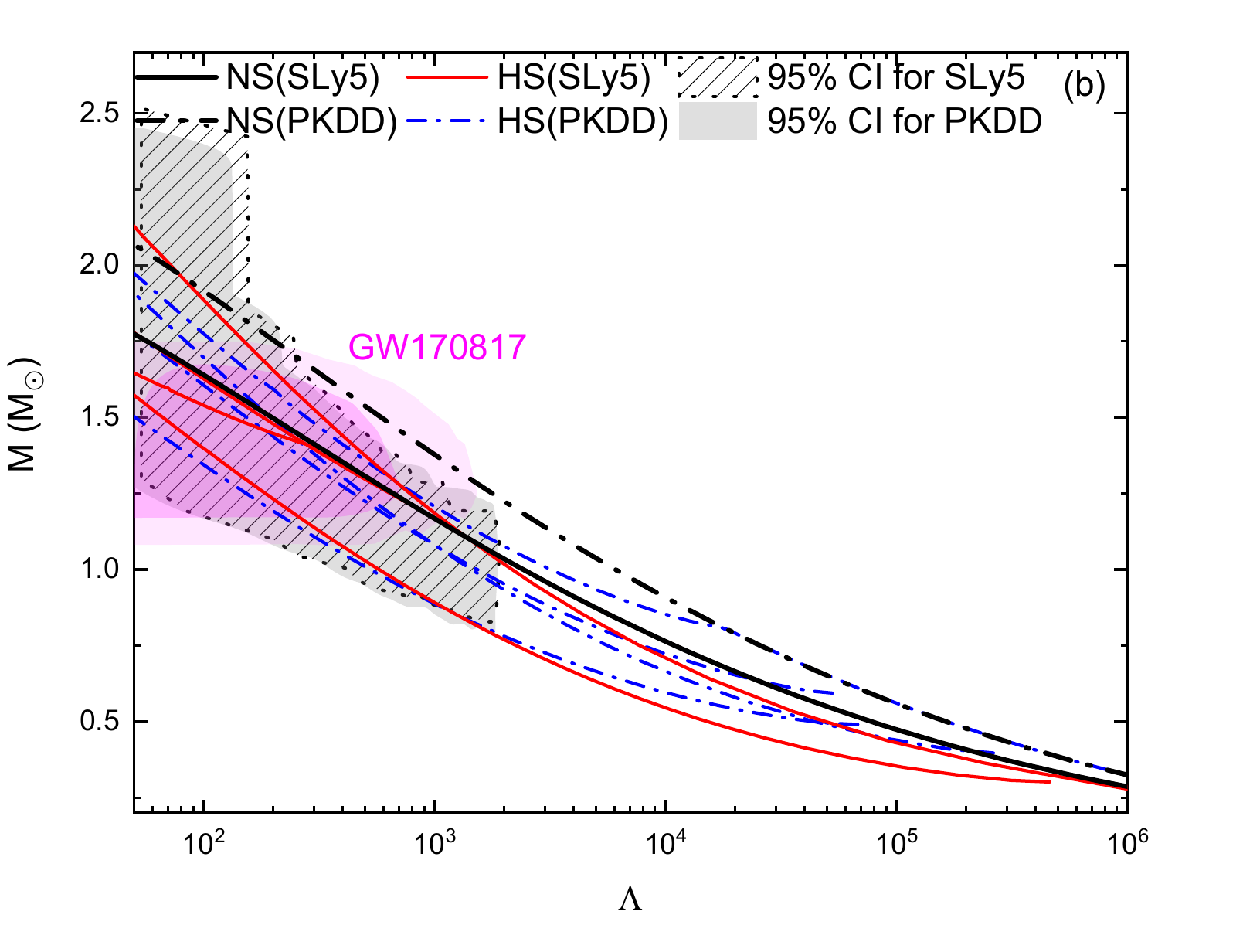}
\end{subfigure}
\caption{Mass-radii (a) and mass-$\Lambda$ (b) solutions of PKDD and SLy5 EoSs with FOPT, where the contours considered here are presented in a 95\% credible interval. Observational data are also represented at 68\% and 95\% confidence intervals.}
\label{fig:MR}
\end{figure}

We now represent the mass-radius ($M-R$) and mass-tidal deformability ($M-\Lambda$) diagrams in Fig.~\ref{fig:MR} for the same cases as detailed in Fig.~\ref{fig:EoSParameters}. Observational contours are also shown in the colored area in Fig.~\ref{fig:MR}(a): the observed mass from J0348, the NICER contours for J0030+0451 and J0740+6620 pulsars. In Fig.~\ref{fig:MR}(b), we show the contour associated with the GW170817 observation. Two observational contours are given, the innermost ones (deep colors) correspond to a 68\% confidence interval, while the outermost ones (lighter colors) correspond to a 95\%. The nucleon EoSs are represented with the thick black lines (solid for SLy5 and dot-dashed for PKDD), and the blue and red curves are samples of hybrid EoS (the same as the ones shown in Fig.~\ref{fig:EoSParameters}). We observe that the two nuclear EoSs are inside the 68\% confidence level drawn for NICER observation for PSR J0030+0451, and only PKDD is compatible with MSP J0740+6620 at 68\% confidence interval, while only SLy5 is inside the 68\% confidence interval for GW170817. 
In other words, PKDD is in better agreement with NICER data, while SLy5 is in better agreement with GW170817. This observation reflects the existence of a tension between the data and the models, since data are well reproduced by different models. This tension was discussed in our previous work~\cite{Guven:2023}, where it was shown that the FOPT is a way to solve it.

In the case of the hybrid EoSs (blue and red curves in Fig.~\ref{fig:MR}), the agreement between the model and the data is better: some samples of hybrid EoS based on both SLy5 and PKDD are within 68\% confidence contours of the NICER observations and the tidal deformability of GW170817. Note, however, that the nature of the FOPT is different for SLy5 and PKDD nuclear EoS: For SLy5, the FOPT should produce a stiff hybrid EoS to be compatible with NICER measurements, while for PKDD, the softer EoS produced by the FOPT approach remains compatible with NICER measurements. The stiff hybrid EoS based on SLy5 is still compatible with the 68\% confidence interval of GW170817, as shown in
Fig.~\ref{fig:MR} (b), while the soft hybrid EoS based on PKDD makes it now compatible with GW170817 at 68\% confidence interval.

In conclusion to this discussion, one can state that the transition to QM is a way to solve the tension between the nuclear EoS on one hand, and the NICER and GW data on the other hand. The nature of the hybrid EoS, however, depends on the stiffness of the nuclear EoS, i.e., on the nuclear symmetry energy. 
Before discussing the role of stiffness and its implication of the nuclear symmetry energy on CSs, 
we first present the uncertainty quantification of the transition to QM and the implementation of nuclear and astrophysical constraints.

\section{Uncertainty quantification and constraints on the quark matter phase}
\label{sec:UQ}

The quantification of the uncertainty of the parameters governing the transition to QM and the implementation of the nuclear and astrophysical constraints are performed within the Bayesian approach.

\subsection{Bayesian approach}
\label{sec:BA}

The Bayesian approach allows a statistical exploration of the uncertainties in the model parameters as well as the limitations in this exploration originating from a set of constraints. 

The core of the Bayesian analysis lies in Bayes' theorem, expressing the probability associated with a given model, represented here by its parameters $\{a_i\}$, to reproduce a set of data. The posterior probability distribution function (PDF) $P(\{a_i\} \mid \textrm{data})$ is obtained as~\cite{Sivia:2006}:
\begin{equation}
\label{e21}
 P(\{a_i\} \mid \textrm{data})\sim P(\textrm{data}\mid \{a_i\})\times P(\{a_i\}),
\end{equation}
where
$P(\textrm{data}\mid \{a_i\})$ is the likelihood probability determined from the comparison between the model and the measurements, and $P(\{a_i\})$ is the prior which represents the \textsl{a priori} knowledge on the model parameters. 

The marginal one- and two-parameter probabilities are defined as~\cite{Sivia:2006}:
\begin{eqnarray}
P(a_j \mid \textrm{data}) &=&\Bigg\{ \prod_{ \substack{i=1\\ i\neq j}}^{N}  \int da_i \Bigg\}  P(\{a_i\} \mid \textrm{data})\, ,  \label{eq:e22a} \\
P(a_j,a_k \mid \textrm{data}) &=& \Bigg\{ \prod_{ \substack{i=1\\ i\neq j,k }}^{N}  \int da_i \Bigg\}  P(\{a_i\} \mid \textrm{data})\, ,  \label{eq:e22b}
\end{eqnarray}
where $N$ is the number of model parameters which are varied for the uncertainty quantification. These marginal probabilities also represent the posterior PDF for one parameter~\eqref{eq:e22a}, and the correlation between two model parameters~\eqref{eq:e22b}.

The likelihood probability defines the ability of the model to reproduce the data and contains limitations in the model parameter space due to the constraints. It is expressed as:
\begin{equation}
\label{e23}
P(\textrm{data}\mid \{a_i\})=w_\textrm{filter} \times p_{\tilde{\Lambda}}\, ,
\end{equation}
where $w_\textrm{filter}(\{a_i\})$ is a pass-band type filter reflecting the constraint limitations, and the probability $p_{\tilde{\Lambda}}$ represents the ability of the model to describe the tidal deformability from the GW170817 event~\cite{Abbott:2019}. These two terms can be expressed as the following constraints:
\begin{itemize}
\item[(C1)] Each viable EoS shall satisfy causality, stability, and positiveness of the symmetry energy, see Ref.~\cite{Margueron:2018b}, and the associated $M_\tov$ should be larger than the largest observed mass $M_{\max}^\mathrm{obs}\approx 2M_\odot$~\cite{Antoniadis:2013}. These constraints shall be satisfied by all EoS up to the density corresponding to the maximum density of the stable branch, denoted $n_\tov$.
\item[(C2)] The probability $p_{\tilde{\Lambda}}$ in Eq.~\eqref{e23} is defined in the following way:
\begin{equation}
\label{e24}
p_{\tilde{\Lambda}} = \frac 1 {N} \sum_{k=1}^{M} p_{\tilde{\Lambda}}^k \, ,
\end{equation}
where the index $k$ runs over a set of binary configurations $\{M_{1,k},\Lambda_{1,k},M_{2,k},\Lambda_{2,k}\}$ for a given EoS (we consider six masses $M_i$) and respecting $0.7 \leq q \leq 1$ and $M_\text{chirp}=1.1975^{+0.0001}_{-0.0001} M_\odot$ for GW170817. In other words, to be compatible with GW170817, the masses $M_1$ and $M_2$ should be distributed between 1.1 and 1.6$M_\odot$. Large masses $M_1>1.6M_\odot$ CSs exist in nature (since $\sim 2M_\odot$ CSs have been detected), but they are not constrained by GW170817. The internal probability $p_{\tilde{\Lambda}}^k$ is obtained by comparing the model prediction and the measurement for GW170817.
\end{itemize}

Another choice, different from (C2), to calculate $p_{\tilde{\Lambda}}$ could be to assign to $p_{\tilde{\Lambda}}$ the maximum probability obtained for the effective tidal deformability $\tilde{\Lambda}$, $p_{\tilde{\Lambda}} = \max_{k} p_{\tilde{\Lambda}}^k$. In the present analysis, we do not explore this other choice to determine $p_{\tilde{\Lambda}}$. It would, however, be interesting to explore the role of the different prescriptions for $p_{\tilde{\Lambda}}$ on the results.

There are three independent GW analyses of GW170817 providing different measurements for $\tilde{\Lambda}$: i) the tidal deformability from Ref.~\cite{Coughlin:2019} includes 
GW, Electromagnetic (EM), and Gamma Ray Burst (GRB) signal, and predict a single peak distribution located around $\tilde{\Lambda}_\textrm{max}\approx 600$; ii) The tidal deformability from Ref.~\cite{De:2018} gives also a single peak around $\tilde{\Lambda}_\textrm{max}\approx 200$; and iii) the tidal deformability of suggested by the LIGO-Virgo collaboration in Ref.~\cite{Abbott:2019}, predicting at least two peaks: the largest one is peaked around $\tilde{\Lambda}_\textrm{max}^1 \approx 180$ and the smaller one is around $\tilde{\Lambda}_\textrm{max}^2\approx 550$, see Fig.~\ref{fig:Lambda} for instance. Since the latter case~\cite{Abbott:2019} corresponds to a middle ground between the analysis of Ref.~\cite{Coughlin:2019} and the one of Ref.~\cite{De:2018}, we choose it as the reference measurement for the tidal deformability of GW170817.

It should be noted that the measurements from NICER are not included in our Bayesian approach. The reason is that different analyses come to different conclusions for a given source. In addition, the sources to consider are still different for different teams. We have therefore decided to compare our predictions in the $M$-$R$ with the NICER measurements, instead of employing them in the EoS selection. There is therefore no constraint from NICER in our approach.

\begin{table}
\centering
\tabcolsep=0.2cm
\def\arraystretch{1.5}
\begin{tabular}{lccc}
\hline\hline
Parameters & Initial Value  & Final Value  & Step \\
\hline
$p_\fopt$ (MeV~fm$^{-3}$) & 6   & 500   & 1 \\
$\mu^*_\QM$ (MeV)         & 950 & 1200  & 2 \\
$\alpha_\fopt$            & 1/3 & 1 & 1/3 \\
\hline\hline
\end{tabular}
\caption{Boundaries and steps in the exploration of FOPT parameters, considering a uniform grid in the parameter space.}
\label{tab:priors}
\end{table}

Let us now discuss the priors. We consider two typical representations of the symmetry energy produced by the soft SLy5~\cite{Chabanat:1998} model and the stiff PKDD~\cite{Long:2004} one. In addition, we vary the FOPT parameters as described in Table~\ref{tab:priors}, guided by the vanishing of the pressure posterior probability at the boundary, and by color-superconducting QM calculations~\cite{Agrawal:2010}.  The sound speed parameter $\alpha_\fopt$ is fixed to three possible values: $\alpha_\fopt=1/3$ (conformal limit), $2/3$ and $1$ (causal limit). We generate about $374\ 000$ priors in total, giving about 187,000 samples for each nuclear EoS. Note that these priors are an extension of the ones considered in our previous work~\cite{Guven:2023}, considering the variation of the chemical potential at zero QM pressure $\mu^*_\QM$. 

We have verified that for these priors, it is possible to obtain CS masses within the interval $[0.5 M_\odot, 1.6 M_\odot]$, as expected to reproduce the GW170817 observation. Note that the maximum mass obtained is well above the observational limit of $2M_\odot$, see Fig.~\ref{fig:MR}.

\subsection{Neutron versus hybrid stars}

In the following, the parameters of the FOPT are varied within the priors, and the conditions (C1) and (C2) are imposed in the Bayesian framework. As a result, the properties of nuclear and hybrid EoS, as well as NS and HS, are given by the grey bands in Figs.~\ref{fig:EoSParameters} and \ref{fig:MR}.

We observe that the energy density varies inside $[200, 800]$~MeV~fm$^{-3}$ in Fig.~\ref{fig:EoSParameters}(a). The latent heat $\Delta \varepsilon_\fopt$ lies inside the following interval $[15,500]$~MeV~fm$^{-3}$ for SLy5 and $[125,500]$~MeV~fm$^{-3}$ for PKDD. The large values for the latent heat for PKDD turn the stiff PKDD nuclear EoS into a softer hybrid EoS compatible with GW170817 observations. For SLy5, the soft nuclear EoS is turned into a hybrid EoS, which can be either soft or stiff, maximally exploring the observational uncertainties.

In terms of particle density, see Fig.~\ref{fig:EoSParameters}(b), the FOPT goes from $n_\fopt\gtrapprox n_\sat$ up to $\approx 4n_\sat$. The low values for $n_\fopt$ are correlated with the large values for the sound speed $c_{s,\fopt}$ that we explore. Our prediction is compatible with Ref.~\cite{Xie:2021}, which predicts QM to appear, on average, for densities around $1.5 n_\sat$. Other analyses predict that QM appear above $3n_\sat$~\cite{Pfaff:2022,Ayriyan:2019} or above $1.6$M$_\odot$~\cite{Parisi:2021}, but, in these studies, high values for the sound speed are excluded. Let us mention that the FOPT model masquerades repulsive EoS as the one predicted by QycM for low values of $n_\fopt$, see Ref.~\cite{Somasundaram:2022b}. Therefore, the low values for $n_\fopt$ that we explore can also be interpreted as a signature of the onset of QycM. In this case, the HS is similar to a quarkyonic star.

For the chemical potential $\mu$, see Fig.~\ref{fig:EoSParameters}(c), note that the hybrid EoSs based on PKDD generate a lower value for the chemical potential (at constant $p$) compared to the nuclear EoS. This can be easily understood since the GW170817 observation requires a softening of the stiff PKDD nuclear EoS, see Fig.~\ref{fig:MR} (b) and related discussions. The softening is performed by lowering the slope of the energy density, i.e., the chemical potential. For the soft nuclear models, e.g. SLy5, the hybrid EoSs can be softer or stiffer than the nuclear EoS and still reproduce the GW170817 observation.

We now discuss Fig.~\ref{fig:MR} (a) and (b), where the grey areas represent the mass-radius ($M-R$) and mass-tidal deformability ($M-\Lambda$) region explored by our statistical approach, considering the constraints (C1) and (C2) (hashed area for SLy5 and plain area for PKDD). While there is a tension to reproduce NICER contours and GW170817 with the two nuclear EoS considered here, the FOPT model can create hybrid models compatible with all data shown in Fig.~\ref{fig:MR}. Some of these hybrid EoSs predict radii which are lower than the measurement from NICER sources, see Fig.~\ref{fig:MR} (a), but they are not excluded since we have not considered NICER measurements in the selection of EoS. Note that hybrid EoSs populate the low region for the tidal deformability that nuclear EoSs do not access, see Fig.~\ref{fig:MR} (b) and Refs.~\cite{Malik2018,Guven:2020,Dinh2021}.

In conclusion, we observe that hybrid EoSs can extend the domain of radii, masses, and tidal deformabilities explored by nuclear EoSs. They are also able to resolve the tension existing between the nuclear EoSs and the astrophysical data.

\subsection{Prediction for GW170817}

\begin{figure}[t]
\centering
\begin{subfigure}{0.5\textwidth}
\includegraphics[width=1\textwidth]{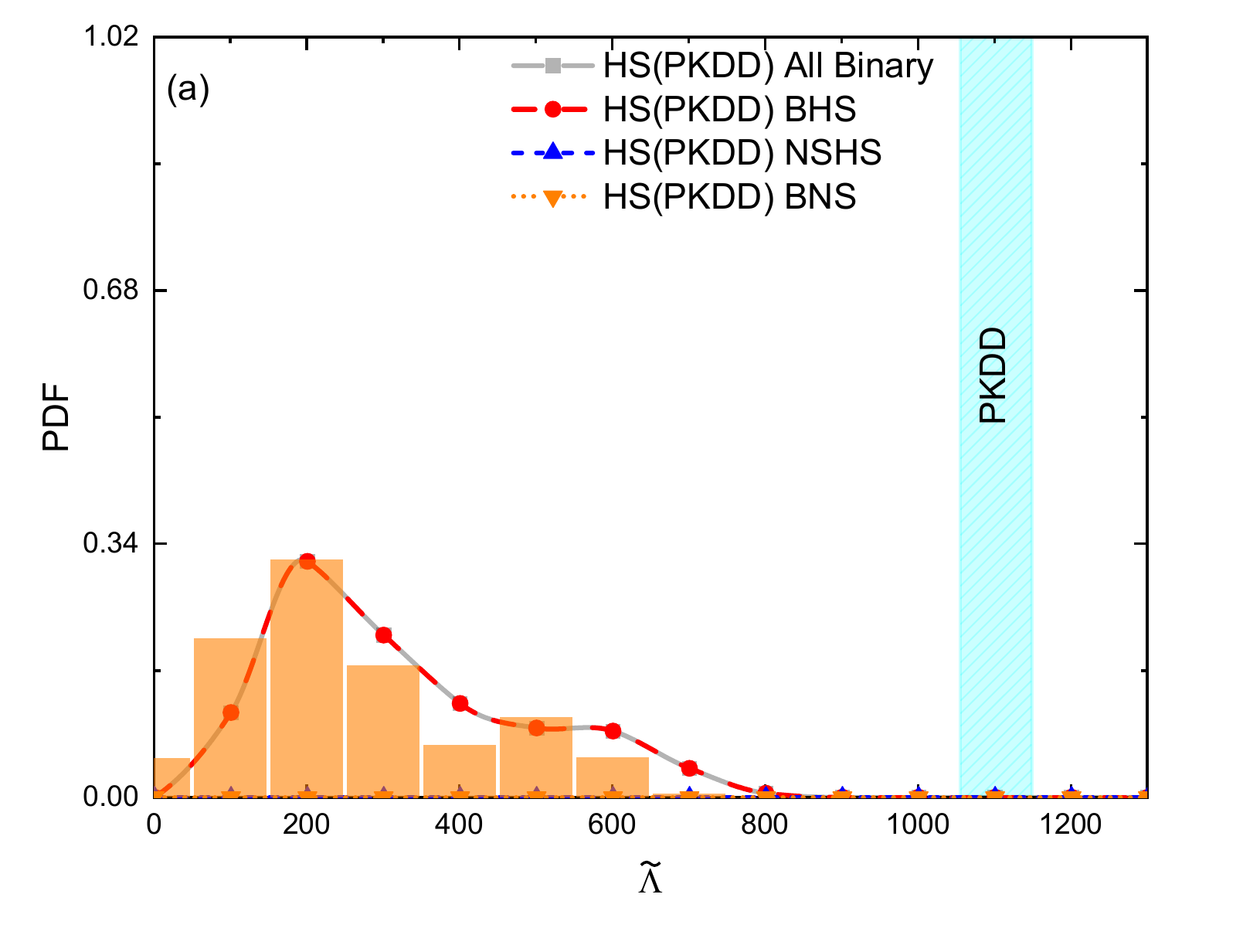}
\end{subfigure}
\begin{subfigure}{0.5\textwidth}
\includegraphics[width=1\textwidth]{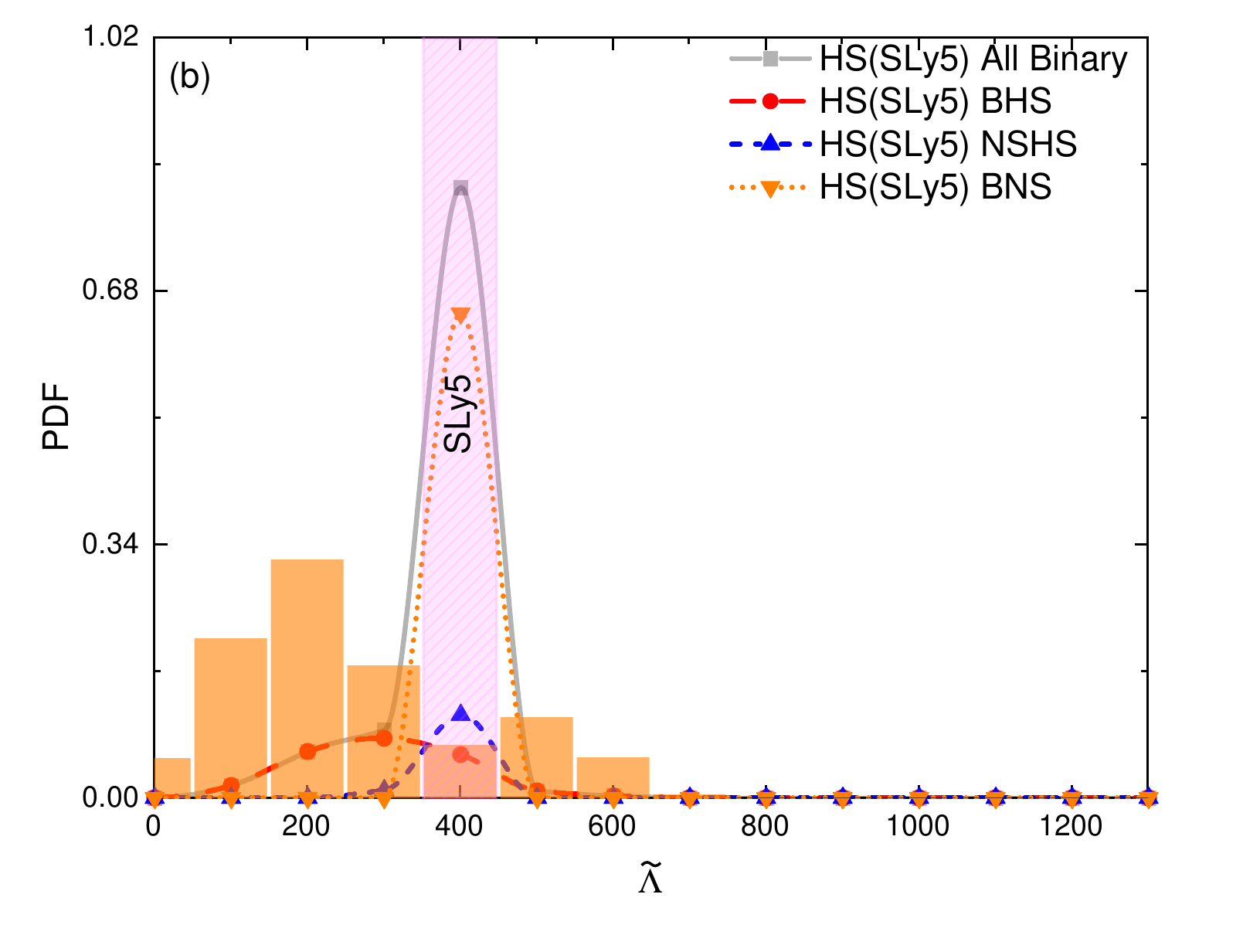}
\end{subfigure}    
\caption{Posterior PDF for $\Tilde{\Lambda}$ with PKDD (a) and SLy5 (b) EoSs with FOPT. The orange band represents tidal deformability of GW170817 from LIGO-Virgo parameter estimation~\cite{Abbott:2019}.}
\label{fig:Lambda}
\end{figure}

The $\tilde{\Lambda}$ measurement determined from the analysis of GW170817~\cite{Abbott:2019} is shown in Fig.~\ref{fig:Lambda} as the orange histogram. It is compared to the distribution of $\tilde{\Lambda}$ from the EoS models considered in our work. To obtain $\tilde{\Lambda}$ from our models, we construct a binary system composed of two CSs with different masses. The mass asymmetry is controlled by the mass ratio $q$, which is varied as suggested by the GW170817 measurement. The binary system can be composed of two NS (BNS), two HS (BHS), or one NS and one HS (NSHS). Note that we name NS a star that remains nuclear up to the GW170817 maximal masses. Above the maximal mass imposed by GW170817, and below the TOV mass, a NS can, however, turn into an HS without impacting the GW170817 measurement. Therefore, it is possible that one NS in a BNS system becomes an HS, provided the FOPT appears above the maximum mass allowed by the GW170817 measurement.

For BNS, the $\tilde{\Lambda}$ prediction is shown as vertical bars in Fig.~\ref{fig:Lambda} (a) for PKDD and (b) for SLy5. 
Note that the width of the bars is arbitrary.
For stiff nuclear EoS, e.g., PKDD shown in Fig.~\ref{fig:Lambda} (a), the prediction for $\tilde{\Lambda}$ is larger than the measurement: we obtain for instance $\tilde{\Lambda}^\mathrm{PKDD}=1100$. For soft EoSs, e.g., SLy5, the value for $\tilde{\Lambda}$ is compatible with the GW170817 measurement, but only with the larger values of the measurement: we find, for instance, $\tilde{\Lambda}^\mathrm{SLy5}=400$. The full uncertainties in the measurement are, however, not explored by the soft EoS exemplified by SLy5 EoS, even after variation of the mass asymmetry parameter $q$. It is indeed a general result that nuclear EoSs predict values for $\tilde{\Lambda}$ that overlap well with the upper part of the GW170817 measurement. It is more difficult for nuclear EoSs to reproduce the lower part of the $\tilde{\Lambda}$ measurement. Future detections of binary systems similar to GW170817 can provide tighter predictions for $\tilde{\Lambda}$ measurement, allowing for an improved selection of EoSs~\cite{Coupechoux:2023}.

We now discuss configurations with one or two HS: NSHS or BHS configurations. Although the NSHS configurations for PKDD give $300\ge\tilde{\Lambda}^\mathrm{PKDD}_\mathrm{NSHS} \ge 1100$, its probability is so small that it is not visible in Fig.~\ref{fig:Lambda} (a). The reason for the small probability is that the NSHS configurations generate too large values for GW170817 $\tilde{\Lambda}$. BHS configurations are preferred, and it is noticeable that the marginalized probability is in very good agreement with the distribution of the measurement (histogram), see Fig.~\ref{fig:Lambda} (a). It should be noted, however, that the FOPT appears at low-density or equivalently low-masses, $M_\mathrm{HS,2}<1.1 M_\odot$ for $q=0.7$. As discussed previously, this fact could also be a signature associated with the onset of QycM.

For BHS and NSHS configurations, where the hybrid EoSs are based on the soft SLy5 nuclear model, a large overlap with the GW170817 measurement is observed in Fig.~\ref{fig:Lambda} (b). Since the posterior PDF is not normalized, the large BNS probability for SLy5 EoS reflects the large number of EoSs for which the FOPT occurs at high density, or in terms of mass $M_\mathrm{HS,1}>1.6 M_\odot$.Hence, a large mass NS can be an HS without impacting GW170817 data, as we already discussed. In this case, the binary system associated with GW170817 is formed of two NS, and the large probability is a statistical bias induced by the 400,000 models in our set. For soft EoS, GW170817 can equally be a BNS, HSNS, or BHS system.

In summary, if the nuclear EoS is stiff, with large values of the symmetry energy as suggested by the analysis of PREX-II data, then a transition to a soft QM EoS is necessary to make it compatible with GW170817. Instead, if the nuclear EoS is soft, compatible with low values of the symmetry energy as suggested by $\chi$EFT models, then GW170817 can equally well be a BNS, NSHS, or BHS binary system. The $\tilde{\Lambda}$ measurement is, however, best described by a BHS configuration based on a stiff nuclear EoS, see Fig.~\ref{fig:Lambda} (a). This conclusion was already obtained in Ref.~\cite{Guven:2023}. More generally, 
we are illustrating here the impact of the nuclear stiffness for the understanding of the nature of the binary system in GW170817.

\subsection{Model parameter PDFs}

We now discuss in more detail the PDF associated with the model parameters ($p_\fopt$, $\mu^*_\QM$, $n_\fopt$), considering different cases for GW170817, such as BNS, NSHS, or BHS.

\begin{figure}[t]
\centering
\begin{subfigure}{0.5\textwidth}
\includegraphics[width=1\textwidth]{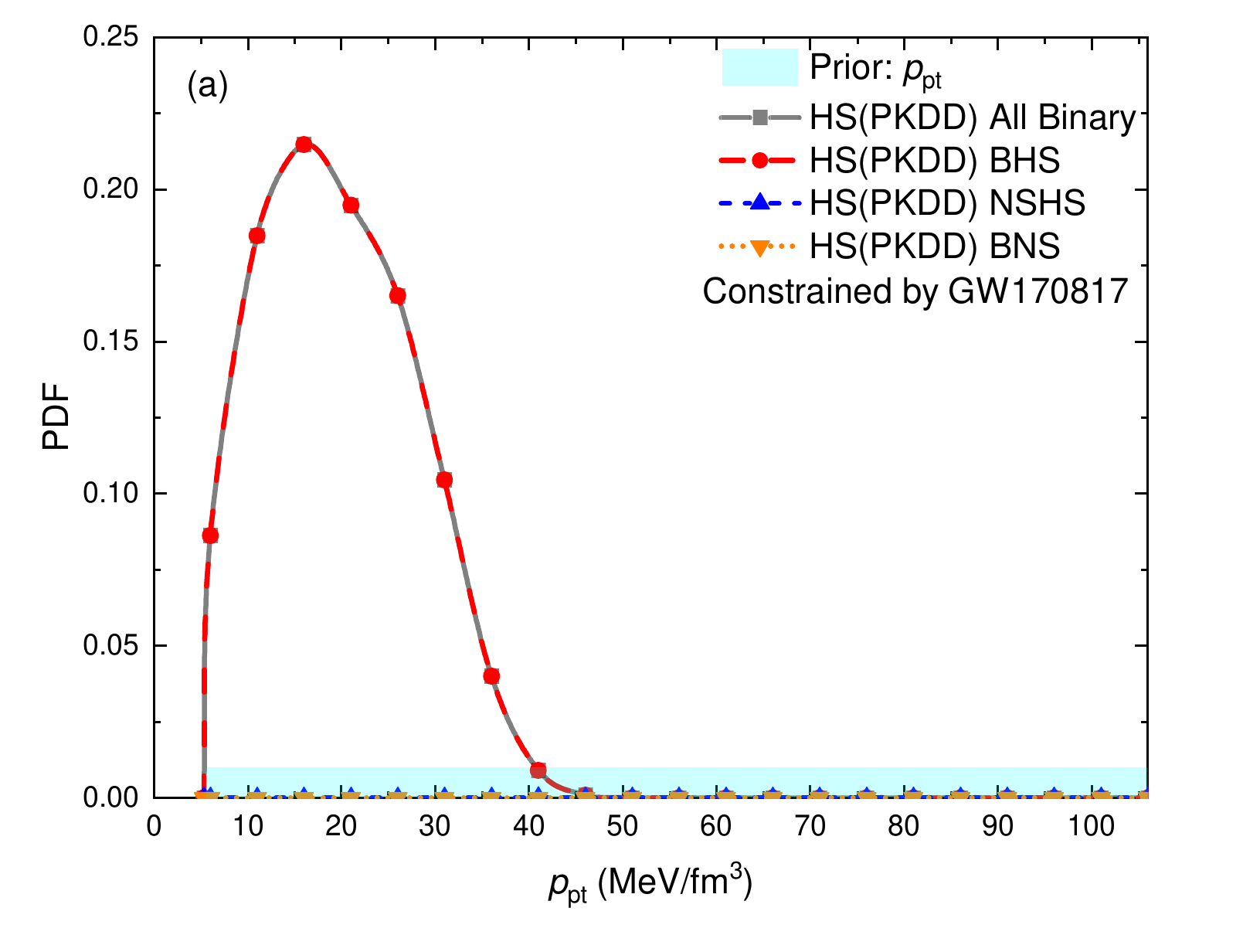}
\end{subfigure}    
\begin{subfigure}{0.5\textwidth}
\includegraphics[width=1\textwidth]{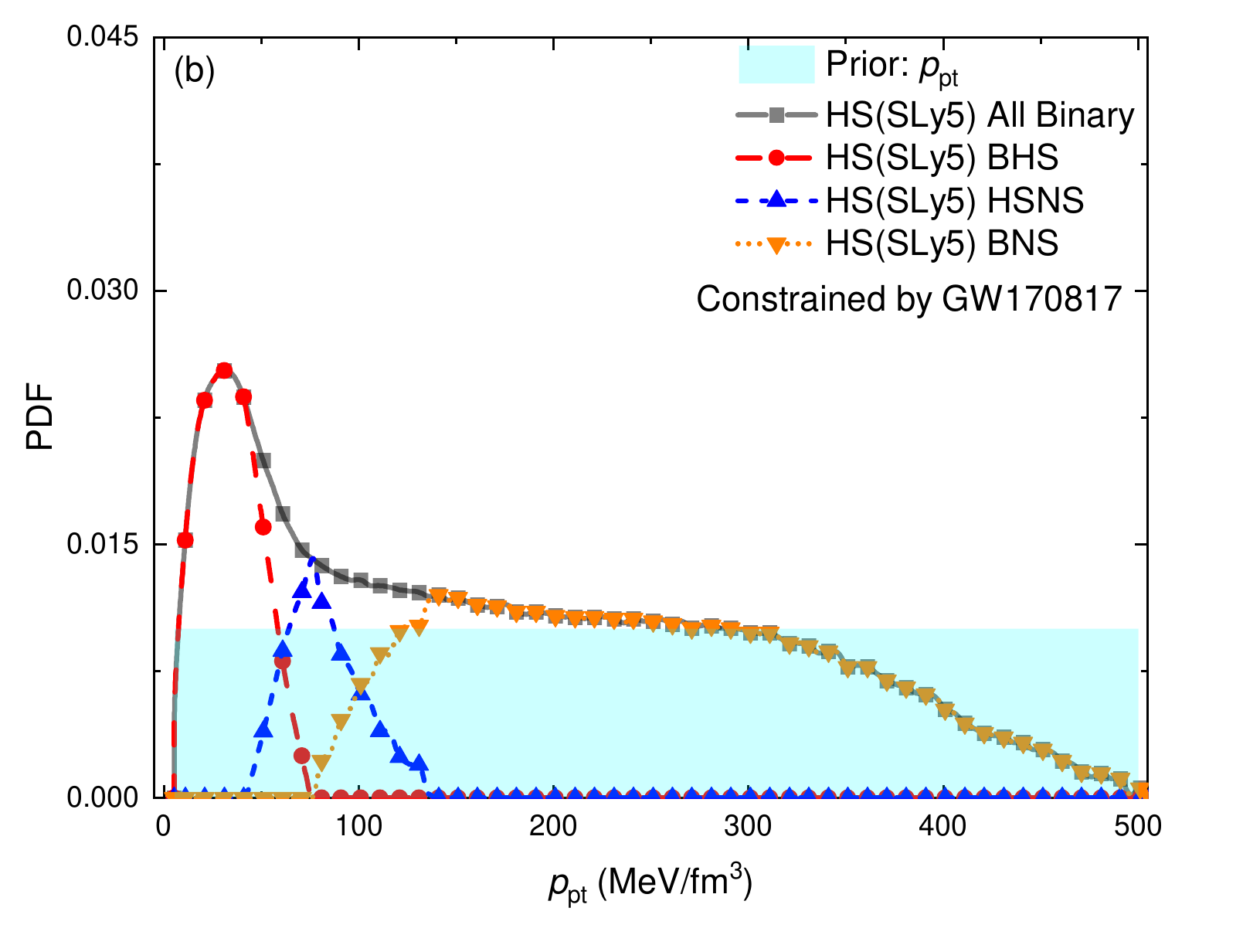}
\end{subfigure}
\caption{Posterior PDF for $p_{\fopt}$ associated with various binary systems (BNS, NSHS, and BHS) and based on PKDD (a) and SLy5  (b) models.} 
\label{fig:Pptall}
\end{figure}

In Fig.~\ref{fig:Pptall}, $p_\fopt$ posterior PDF is shown for binary configurations (BNS, NSHS, and BHS) where the two CSs are obtained fixing the nuclear EoS to be PKDD (a) or SLy5 (b). As expected from the previous discussion, for the EoS based on the nuclear model PKDD, only the BHS configuration is possible.

It is also observed that the BNS configuration is possible only for SLy5 soft nuclear EoS and $p_\fopt\in[100,500]$~MeV~fm$^{-3}$ approximately. The large values for $p_\fopt$ imply that the FOPT occurs at high densities. This configuration remains a BNS for masses compatible with GW170817, but a FOPT can occur for larger masses. Note that for BNS, there is a natural cut-off for large $p_\fopt>500$~MeV~fm$^{-3}$ since above this value, NS are unstable. The mixed NSHS configurations are only possible with SLy5 soft nuclear EoS and $p_\fopt\in[50,100]$~MeV~fm$^{-3}$ approximately (the PDF is peaked at $p_\fopt^\mathrm{SLy5}=83 \pm 20$~MeV~fm$^{-3}$). The small interval for $p_\fopt$ is constrained by the necessity that the FOPT occurs in the mass range compatible with GW170817 (between 1.1 and 1.6 M$_\odot$).

As seen in Fig.~\ref{fig:Pptall}, the BHS configuration is possible for both PKDD and SLy5 nuclear EoS. We have $p_\fopt^\mathrm{SLy5}=34 \pm 16$~MeV~fm$^{-3}$ for PKDD and $p_\fopt^\mathrm{PKDD}=19 \pm 8$~MeV~fm$^{-3}$ for SLy5. For the BHS configuration, there is a large overlap between the $p_\fopt$ PDF. In other words, the BHS configuration is weakly dependent on the nuclear EoS (or on the symmetry energy). The BNS and HSNS configurations require, however, a soft nuclear EoS.

If the nuclear pressure is large enough (higher symmetry energy or stiffer EoS, such as PKDD), 
then NSHS and BNS configurations are excluded from the GW170817 measurement. A better knowledge of the properties of the symmetry energy (stiff or soft) will clarify the nature of the binary system in GW170817.

\begin{figure}
\centering
\begin{subfigure}{0.5\textwidth}
\includegraphics[width=1\textwidth]{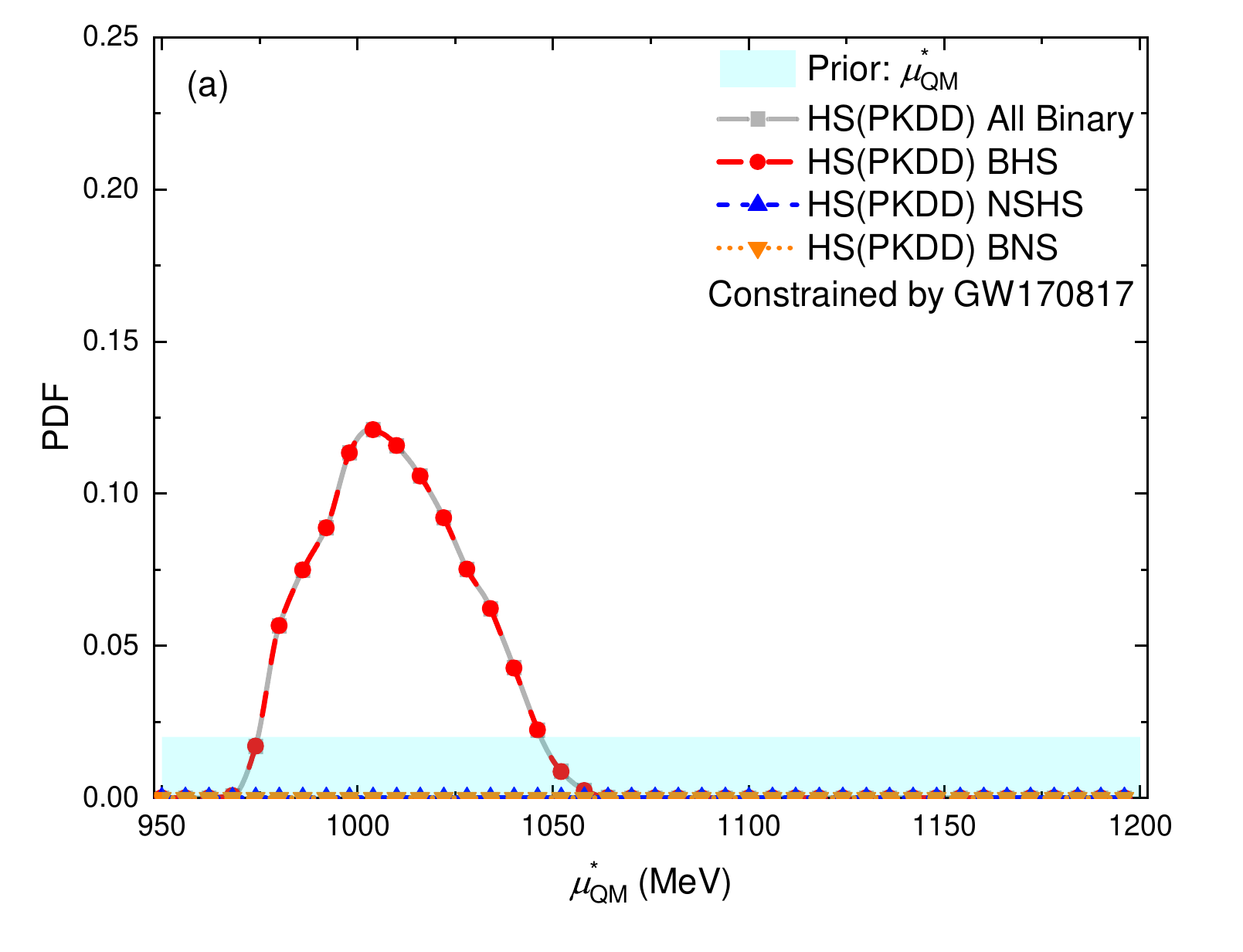}
\end{subfigure}
\begin{subfigure}{0.5\textwidth}
\includegraphics[width=1\textwidth]{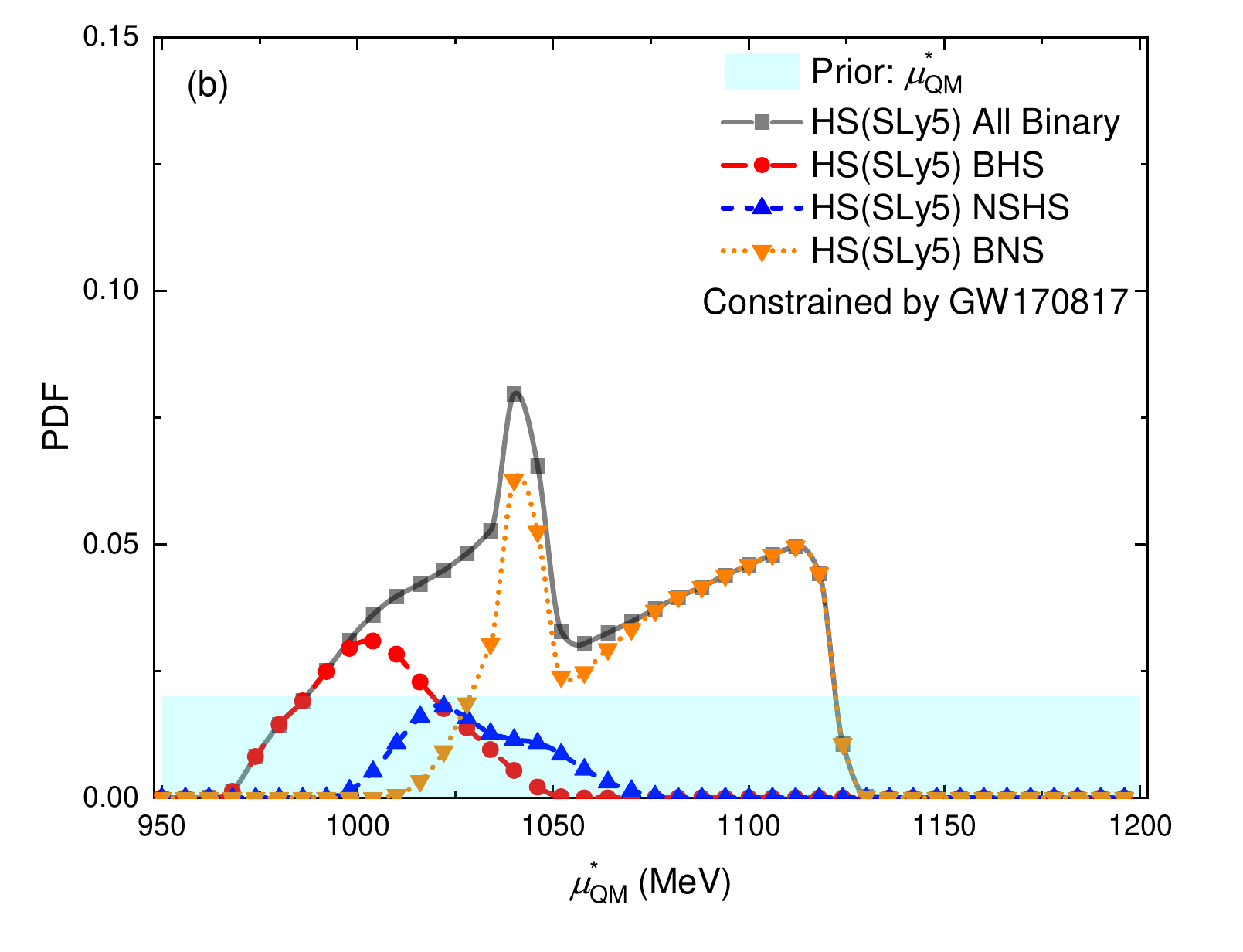}
\end{subfigure}
\caption{Same as Fig.~\ref{fig:Pptall} for $\mu^*_\QM$ PDF.}
\label{fig:mustarall}
\end{figure}

In Fig.~\ref{fig:mustarall} (a) and (b), the $\mu^*_\QM$ PDF associated with various configurations and based on PKDD (a) and SLy5 (b) nuclear EoS is represented. As discussed in Fig.~\ref{fig:Pptall}, 
the BHS configuration is weakly dependent on stiffness
and requires low values for the chemical potential $\mu^*_\QM$: $\mu^{*,\mathrm{PKDD}}_\QM=1009 \pm 18$~MeV and $\mu^{*,\mathrm{SLy5}}_\QM=1005 \pm 17$~MeV. For BHS configurations, the value of the parameter $\mu^*_\QM$ is independent from the symmetry energy.

BNS and NSHS configurations are possible only for the soft nuclear EoS (SLy5). The mixed NSHS configuration compatible with GW170817 imposes the distribution for $\mu^*_\QM$ being in a small interval: between 1000 and 1070 MeV, or $\mu^{*,\mathrm{SLy5}}_\QM=1045 \pm 17$~MeV.
For BNS configurations, the cutoff for large $\mu^*_\QM$ is related to the fact that there are no stable NS if $\mu^*_\QM$ is above 1130~MeV.

\begin{figure}
\centering
\begin{subfigure}{0.5\textwidth}
\includegraphics[width=1\textwidth]{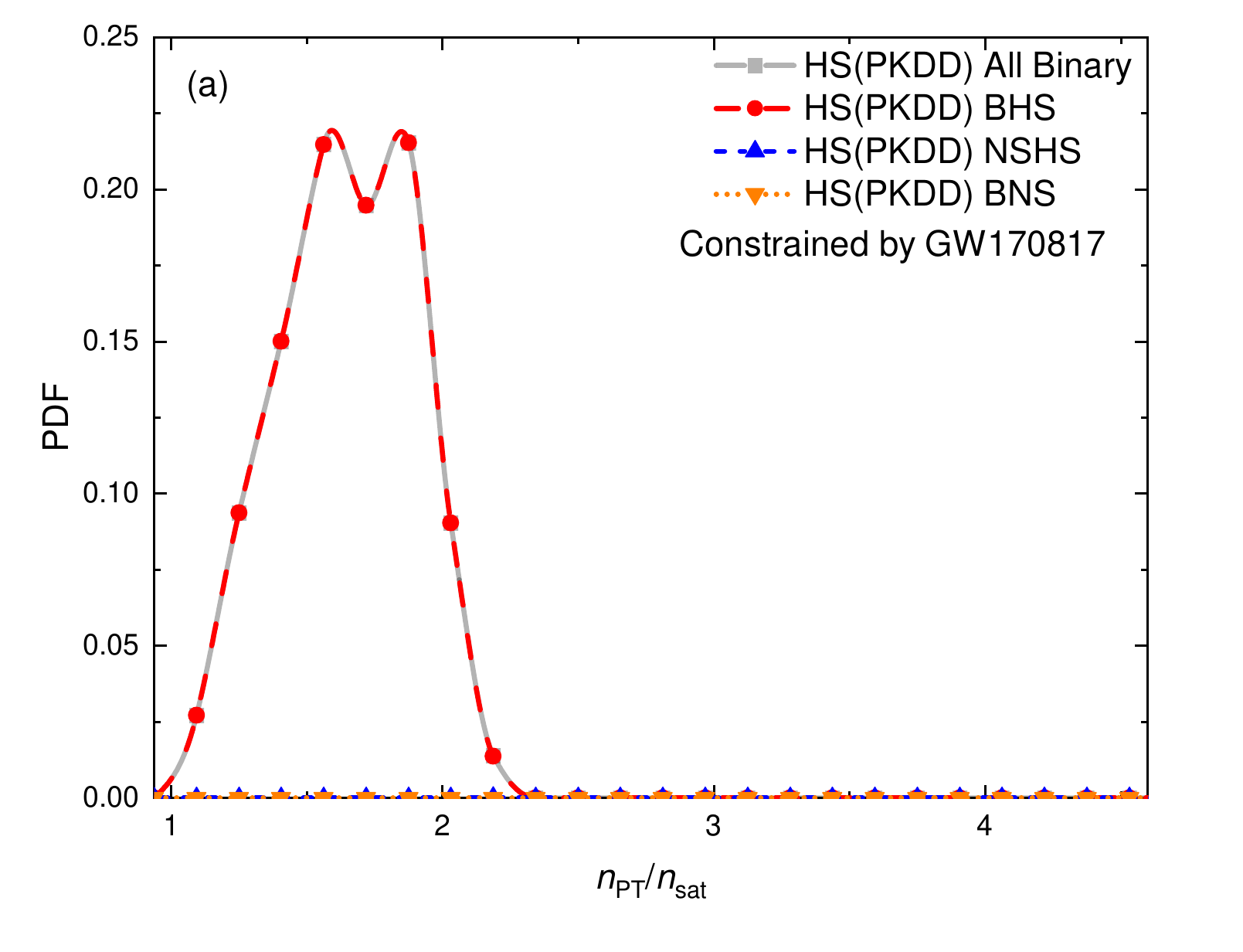}
\end{subfigure}    
\begin{subfigure}{0.5\textwidth}
\includegraphics[width=1\textwidth]{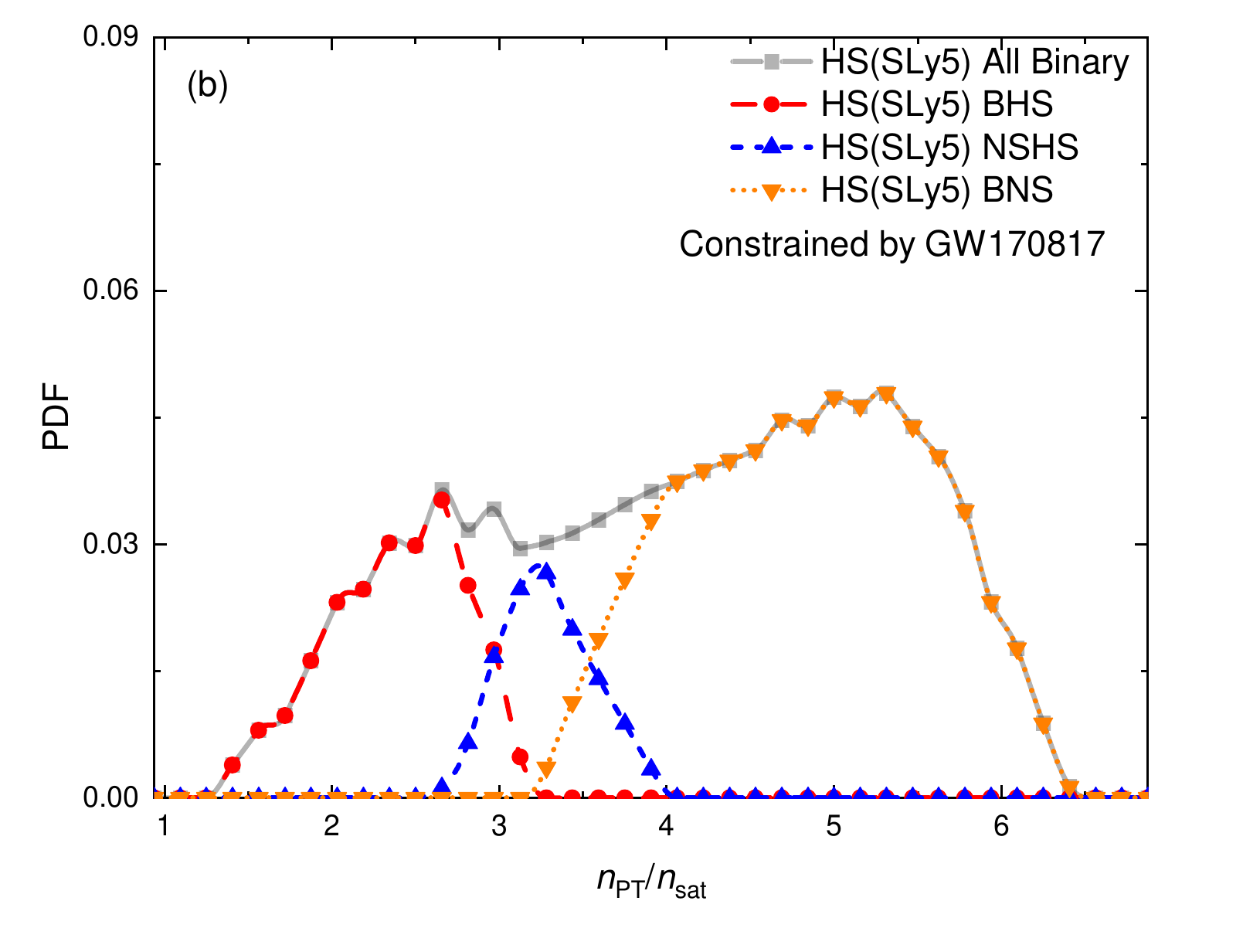}
\end{subfigure}
\caption{Same as Fig.~\ref{fig:Pptall} for $n_\fopt/n_\sat$ PDF.}
\label{fig:ntrall}
\end{figure}

\begin{figure}[t]
\centering
\begin{subfigure}{0.5\textwidth}
\includegraphics[width=1\textwidth]{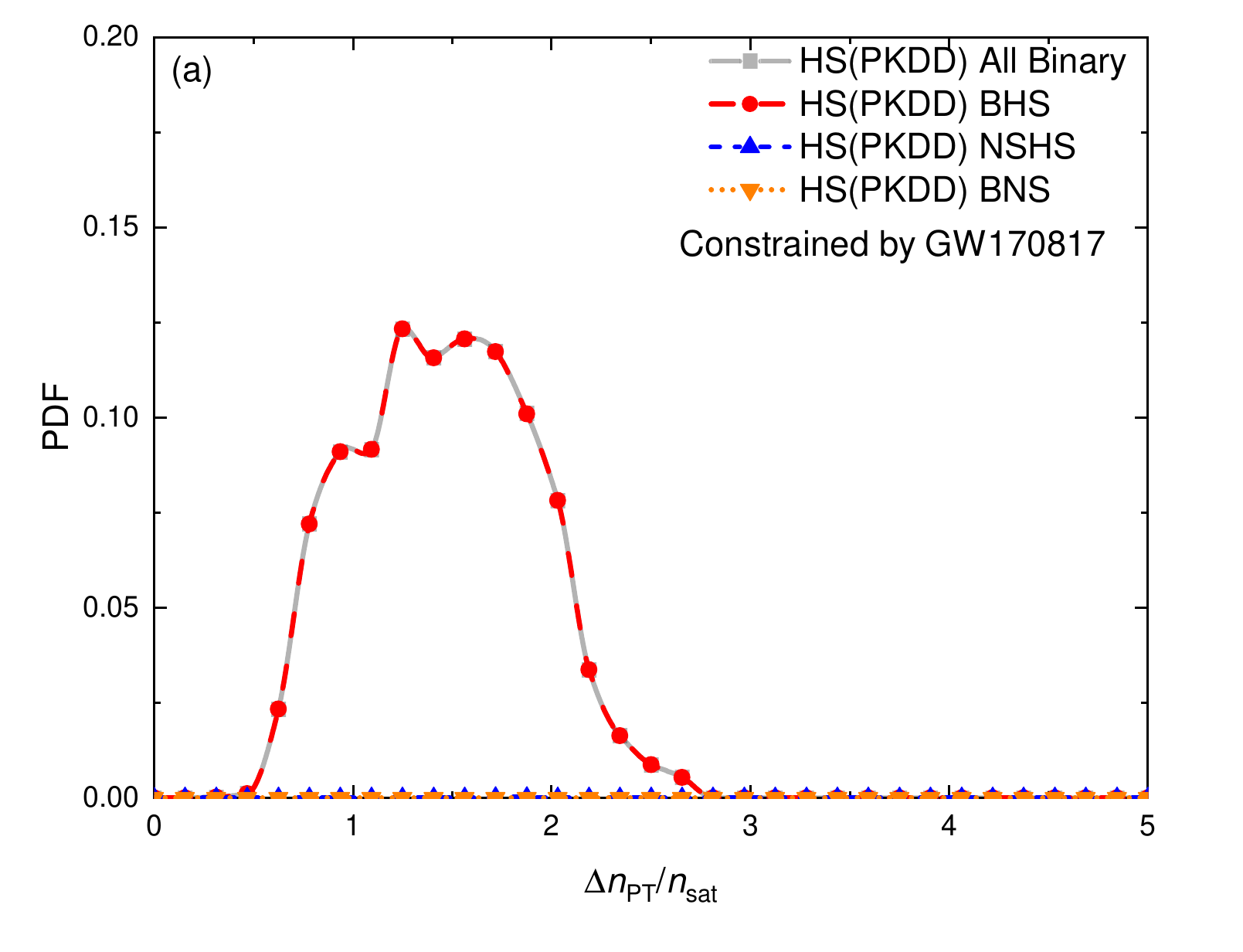}
\end{subfigure}   
\begin{subfigure}{0.5\textwidth}
\includegraphics[width=1\textwidth]{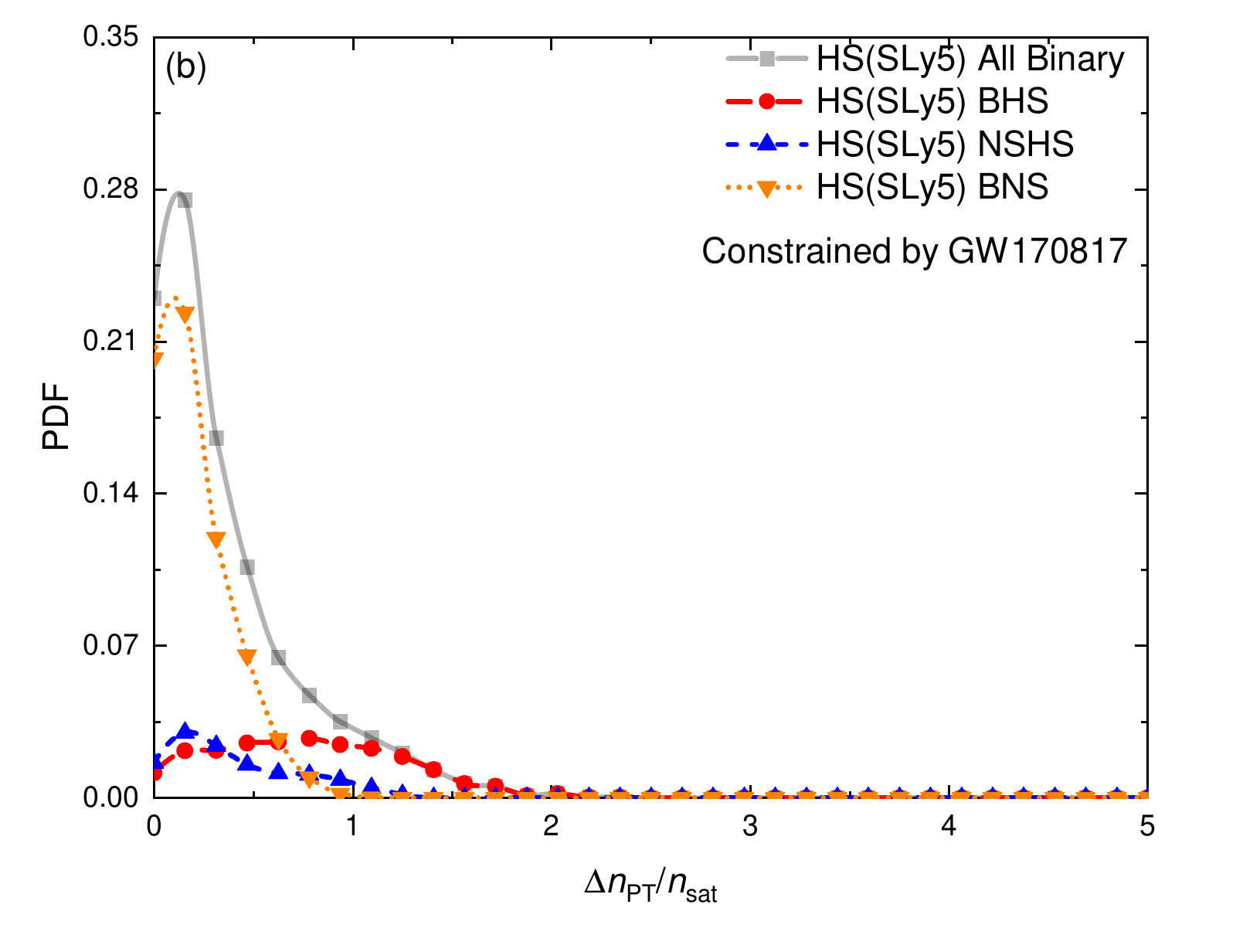}
\end{subfigure}
\caption{Same as Fig.~\ref{fig:Pptall} for $\Delta n_\fopt/n_\sat$ PDF.}
\label{fig:Dn}
\end{figure}

In summary, Fig.~\ref{fig:mustarall} shows that the nature of the QM EoS depends on the stiffness of nuclear EoS as strongly influenced by the nuclear symmetry energy. 
For soft EoS (low values of the symmetry energy), two-flavor color superconducting EoS and color-flavor-locked superconducting EoS are possible (see section II.B for the relation between the $\mu^*_\QM$ value and the flavor-type model), while for stiff EoS (large value of the symmetry energy), only color-flavor-locked superconducting EoS is possible.

We now focus on the density parameters $n_\fopt$ and $\Delta n_\fopt$. Their PDF is displayed in Figs.~\ref{fig:ntrall} and \ref{fig:Dn}.
The lower the FOPT $n_\fopt$, the stiffer the hybrid EoS, due to the constraint (C1), while in general, the larger the density jump $\Delta n_\fopt$ the softer the hybrid EoS. For BHS configurations, soft and stiff nuclear EoS suggest a FOPT in the range of $n_\sat \lessapprox n_\fopt \lessapprox 2$-$3n_\sat$. From the distribution of $n_\fopt$, we obtain $n_\fopt^\mathrm{PKDD}=0.26 \pm 0.04$~fm$^{-3}$ and $n_\fopt^\mathrm{SLy5}=0.38 \pm 0.07$~fm$^{-3}$. 

For a stiff nuclear EoS, such as PKDD, the hybrid EoS is made soft by allowing $\Delta n_\fopt$ to be large: $\Delta n_\fopt\gtrapprox n_\sat$. We obtain $\Delta n_\fopt^\mathrm{PKDD}=0.23 \pm 0.07$~fm$^{-3}$. 

For the soft nuclear EoS, such as SLy5, the hybrid EoS can be soft or stiff since the distribution of $\Delta n_\fopt$ includes values close to zero as well as values as large as $1$-$2n_\sat$, i.e., $\Delta n_\fopt^\mathrm{SLy5}=0.12 \pm 0.07$~fm$^{-3}$.
NSHS configurations are only allowed with soft nuclear EoS, e.g., SLy5 as previously discussed.
The onset density $n_\fopt$ and the density discontinuity distributions are $n_\fopt^\mathrm{SLy5}=0.53 \pm 0.04$~fm$^{-3}$ and $\Delta n_\fopt^\mathrm{SLy5}=0.07 \pm 0.06$~fm$^{-3}$. In this case, note that the FOPT lies in the mass range of GW170817, i.e., for masses between $1.1$ and $1.6M_\odot$, and the hybrid EoS is neither soft nor stiff.

For BNS configurations, the FOPT onset density is large enough to be out of the observation range of GW170817, i.e., $M>1.6M_\odot$. It implies that the onset density distribution is $n_\fopt \leq 0.65$~fm$^{-3}$. The density discontinuity is small, since large discontinuities do not match with the request that the hybrid EoS must support $\sim 2 M_\odot$.

\subsection{Two-parameter correlations} 

We now analyze the correlations among the FOPT parameters $p_\fopt$, $\mu^*_\QM$, $n_\fopt$ and $\Delta n_\fopt$ that fix the FOPT properties. The Pearson coefficients associated with all possible correlations among these FOPT parameters are given in Table~\ref{tab:Corr}. 

\begin{table}
\tabcolsep=0.4cm
\def\arraystretch{1.5}
\begin{tabular}{cccc}
\hline\hline
Correlation & nuclear & \multicolumn{2}{c}{Binary configurations} \\
between & EOS & BHS & NS+(NS or HS)   \\
\hline
$p_{\fopt}$ and $\mu^*_\QM$ &SLy5& 0.91 & 0.77 \\
 &PKDD& 0.96 & 0.89 \\
$p_{\fopt}$ and $n_{\fopt}$ &SLy5& 0.98 & 0.99 \\
 &PKDD& 0.99 & 0.96 \\
$p_{\fopt}$ and $\Delta n_\fopt$&SLy5 &-0.12 & 0.33  \\
&PKDD &0.15 & 0.56   \\
$\mu^*_\QM$ and $n_\fopt$ &SLy5 & 0.89 &0.76\\ 
&PKDD & 0.95 & 0.92\\
$\mu^*_\QM$ and $\Delta n_\fopt$ &SLy5 &0.24 &  0.82 \\
&PKDD &0.41 &  0.85  \\
$\Delta n_\fopt$ and $n_\fopt$ &SLy5 &-0.15  &0.33\\ 
&PKDD & 0.16 & 0.64\\
\hline\hline
\end{tabular}
\caption{Pearson/correlation coefficients associated with all possible correlations among parameters governing the FOPT properties, for BHS and binary systems with at least one NS (NS or HS).}
\label{tab:Corr}
\end{table}

The Pearson/correlation coefficients given in Table~\ref{tab:Corr} for hybrid EoS depend on the nuclear symmetry energy, which is represented by the nuclear EoSs (soft SLy5 or stiff PKDD) given in column two of the table. Since we previously observed differences between BHS configurations on one hand, and BNS and NSHS configurations on the other hand, we decided to group together BNS and NSHS. So two different configurations are investigated: BHS and a binary system with at least one NS, the companion being a NS or a HS: NS+(NS or HS). The largest correlations are obtained for $p_{\fopt}$-$\mu^*_\QM$, $p_{\fopt}$-$n_{\fopt}$ and $\mu^*_\QM$-$n_{\fopt}$ for both BHS and NS+(NS or HS) configurations. The other FOPT parameters are weakly correlated. In the following, we, therefore, investigate further the three cases with the larger correlations.

\begin{figure} 
\centering
\begin{subfigure}{0.5\textwidth}
\includegraphics[width=1\textwidth]{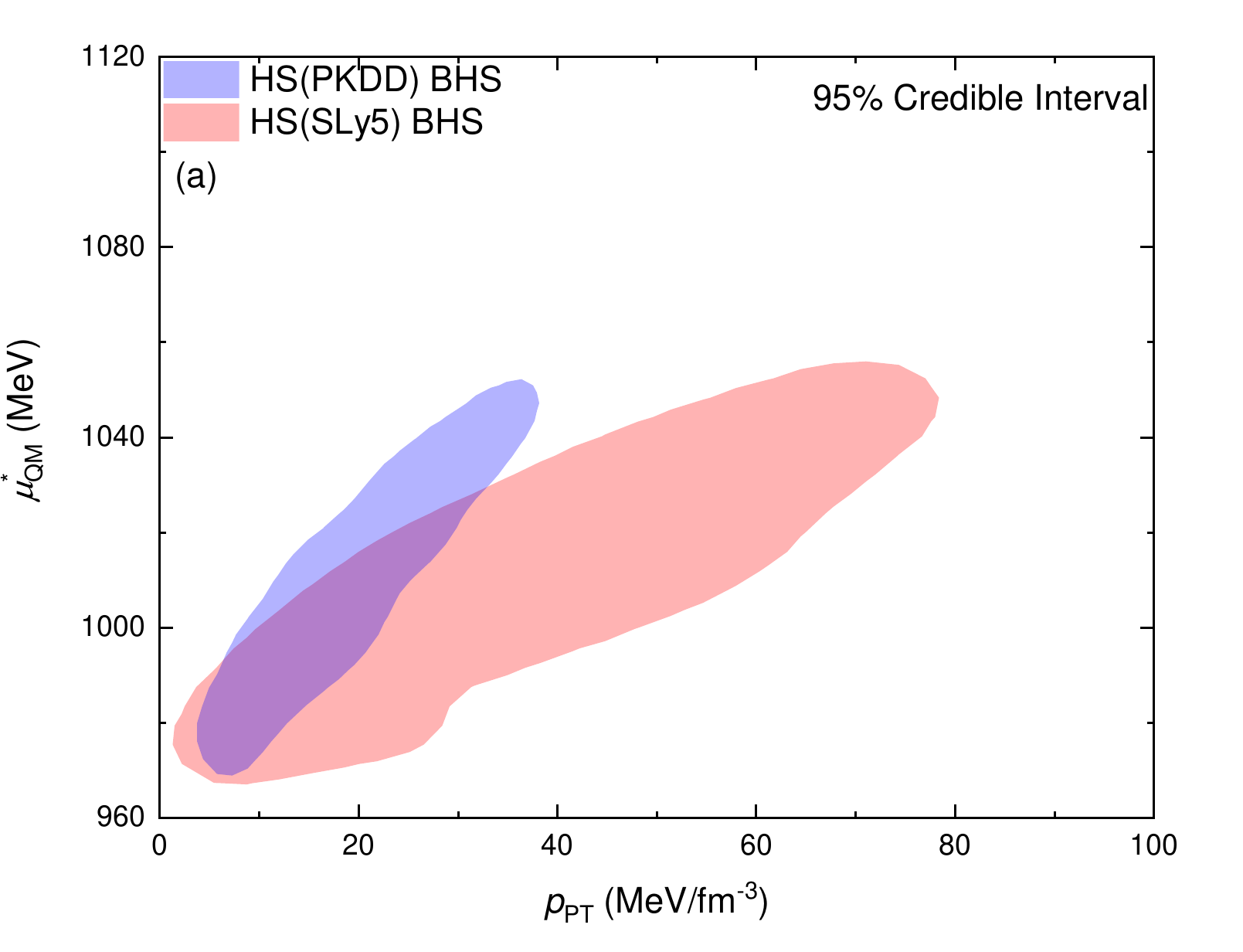}
\end{subfigure}    
\begin{subfigure}{0.5\textwidth}
\includegraphics[width=1\textwidth]{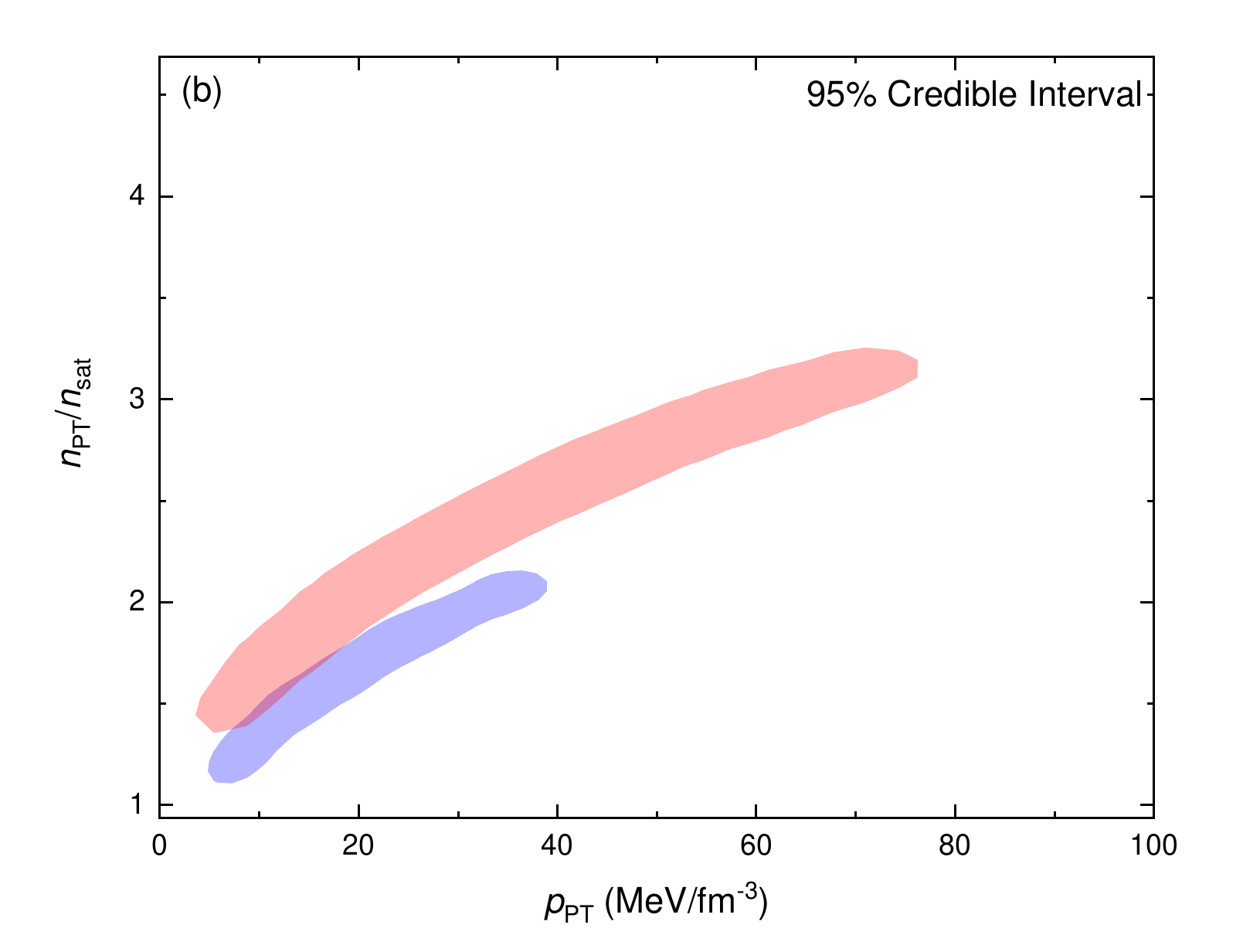}
\end{subfigure}    
\caption{$p_{\fopt}$-$\mu^*_\QM$ (a) and $p_{\fopt}$-$n_\fopt/n_\sat$ (b) contours for 95\% credible interval associated with the BHS configuration and for the soft (SLy5, red) and stiff (PKDD, blue) hybrid EoS.}
\label{fig:Cor:BHS}
\end{figure}

\begin{figure} 
\centering
\begin{subfigure}{0.5\textwidth}
\includegraphics[width=1\textwidth]{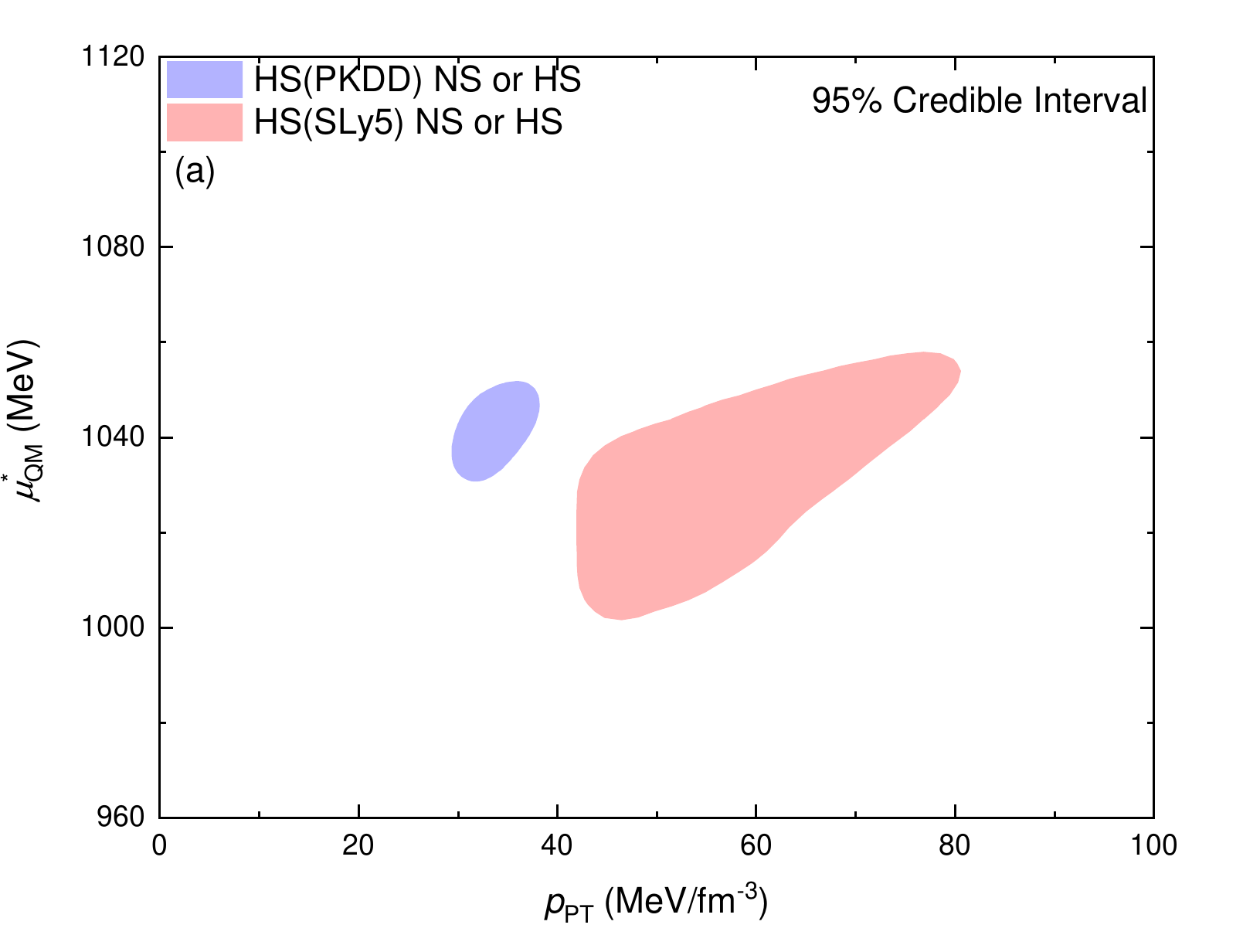}
\end{subfigure}
\begin{subfigure}{0.5\textwidth}
\includegraphics[width=1\textwidth]{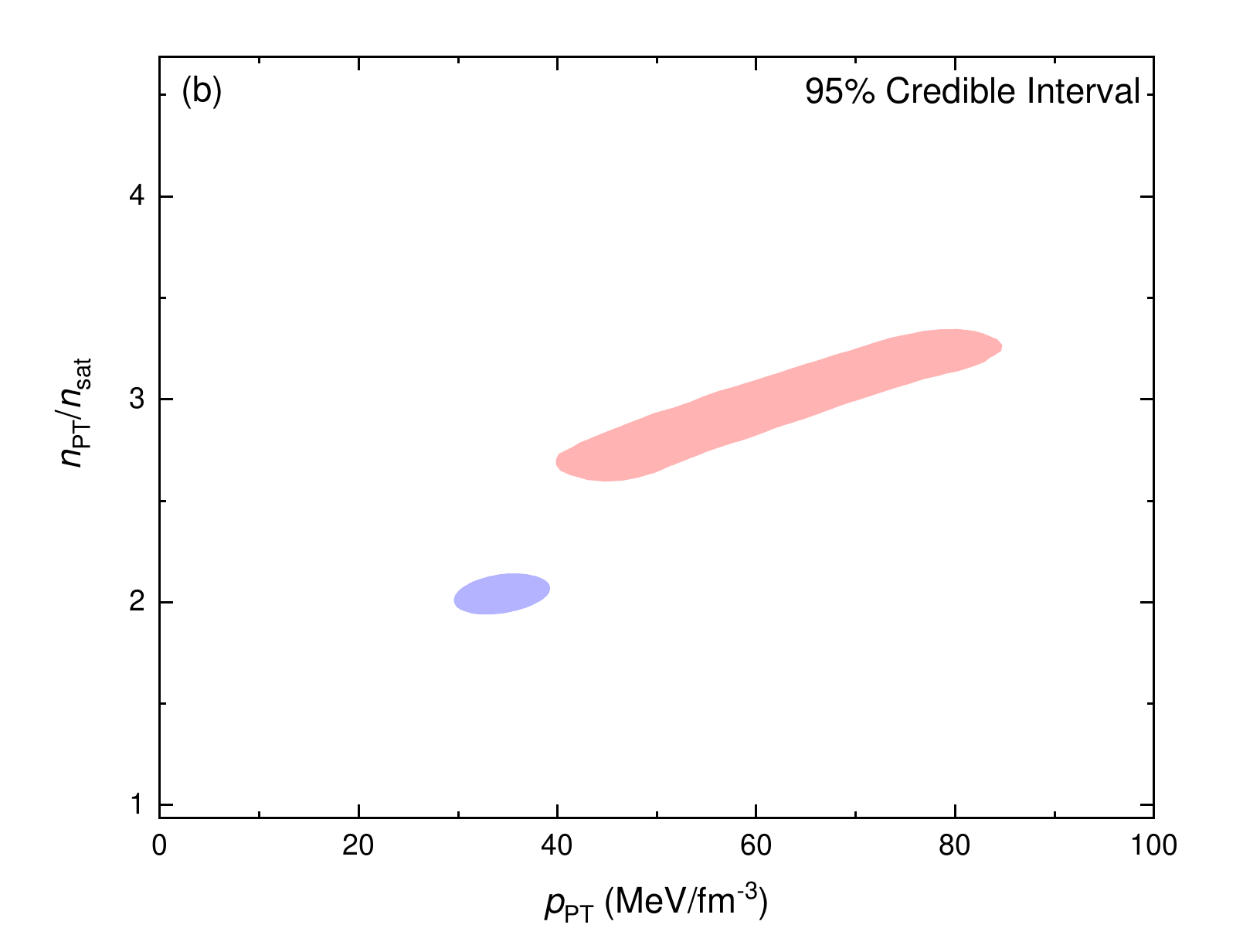}
\end{subfigure}
\caption{$p_{\fopt}$-$\mu^*_\QM$ (a) and $p_{\fopt}$-$n_\fopt/n_\sat$ (b) contours for 95\% credible interval associated with the NS or HS configuration and for the soft (SLy5, red) and stiff (PKDD, blue) hybrid EoS.}
\label{fig:Cor:NSorHS}
\end{figure}

The correlations between $p_{\fopt}$ and $\mu^*_\QM$ and between $p_{\fopt}$ and $n_{\fopt}$ for BHS are shown in Fig.~\ref{fig:Cor:BHS} (a) and (b), where the red region represents the correlation for the hybrid EoS based on the soft SLy5 nuclear model and the blue region for the hybrid EoS based on the stiff PKDD nuclear model. 
There is a clear influence of the nuclear stiffness that is impacted by the nuclear symmetry energy on the correlation, 
the stiffer nuclear EoS privileging larger $\mu^*_\QM$ and smaller $n_\fopt$ for a fixed value of $p_\fopt$ compared to the soft nuclear EoS.
Soft SLy5 nuclear EoS allows a much larger number of EoS since its tidal deformability is closer to that obtained from GW170817. This is also true for the NS+(NS or HS) case in Fig.~\ref{fig:Cor:NSorHS}, where fewer EoS survived the PKDD nuclear model compared to the soft SLy5 nuclear case.
So for a fixed value of $p_\fopt$, stiff nuclear EoS implies lower values for $n_\fopt$ compared to soft nuclear EoS. The chemical potential $\mu^*_\QM$ is larger to stiff nuclear EoS compared to the softer case.

\begin{figure}
\centering
\begin{subfigure}{0.5\textwidth}
\includegraphics[width=1\textwidth]{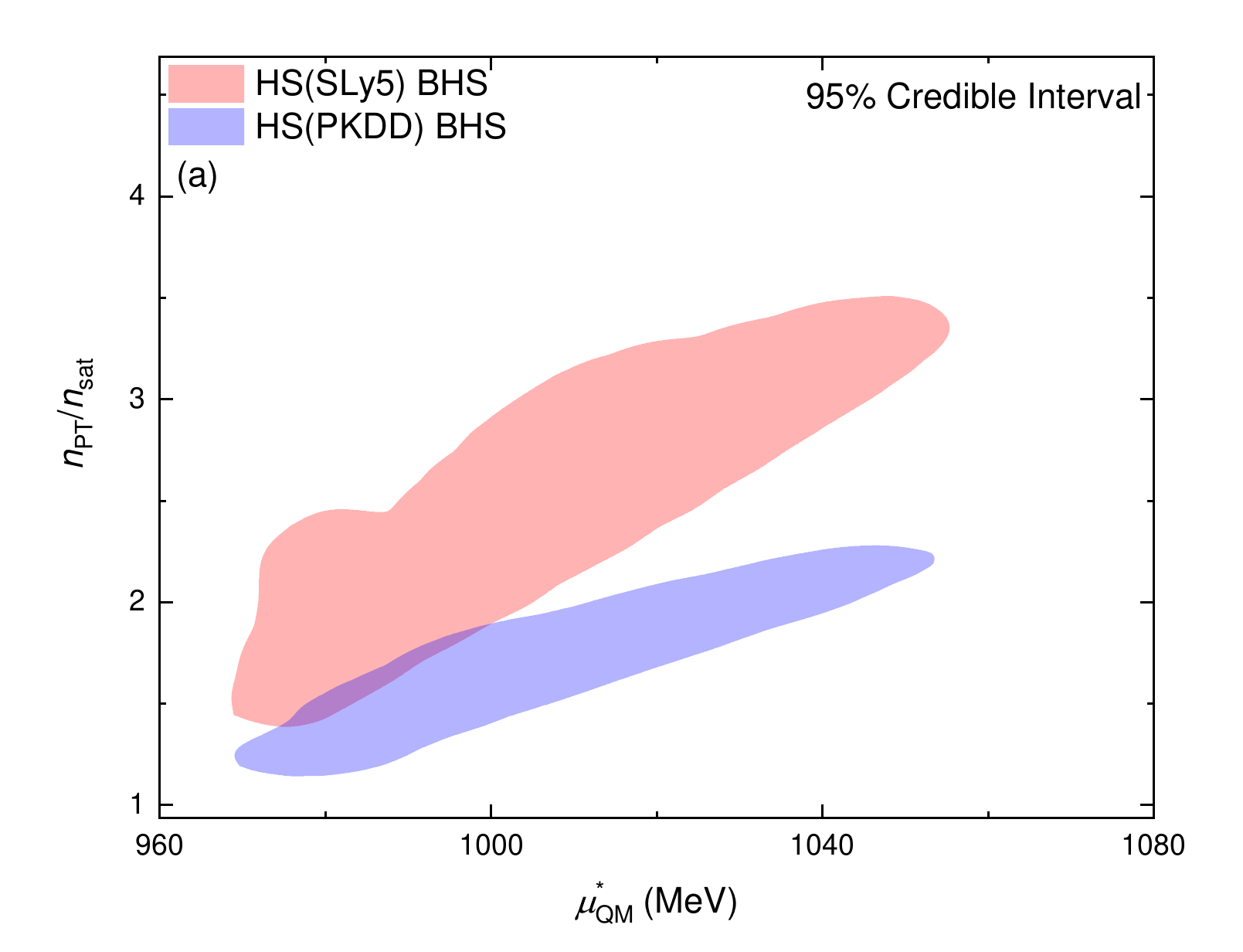}
\end{subfigure}
\begin{subfigure}{0.5\textwidth}
\includegraphics[width=1\textwidth]{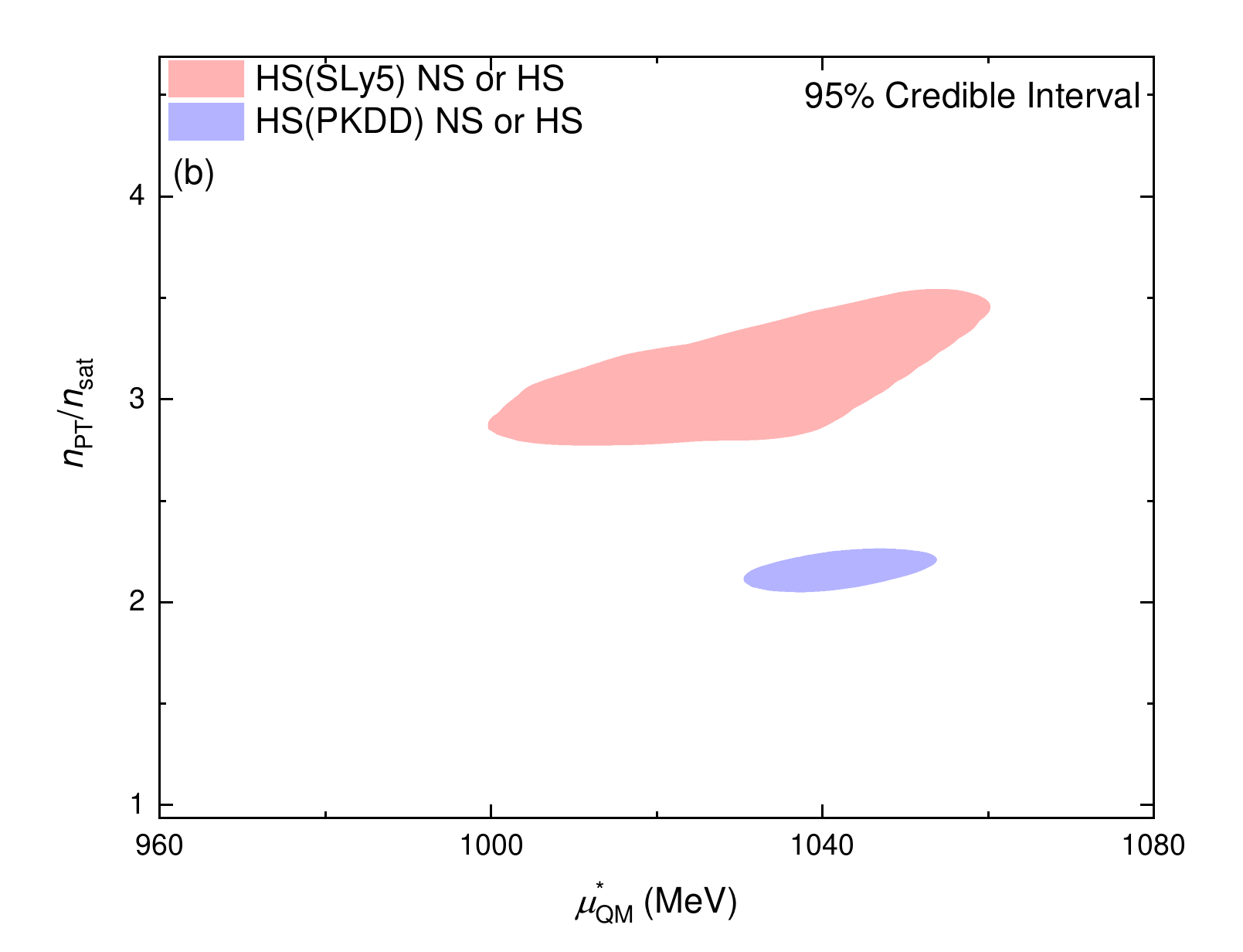}
\end{subfigure}
\caption{$\mu^*_\QM$-$n_\fopt/n_\sat$ contours for 95\% credible interval associated with BHS (a) and the NS or HS (b) configurations for the soft (SLy5, red) and stiff (PKDD, blue) hybrid EoS.}
\label{fig:Cor:Pptntr}
\end{figure}

The correlation of $\mu^*_\QM-n_{\fopt}$ is displayed in Fig.~\ref{fig:Cor:Pptntr}. It shows similar behavior as the two other correlation diagrams shown for $p_{\fopt}$-$\mu^*_\QM$ and $p_{\fopt}$-$n_\fopt$. Since the stiff PKDD favors a low-density FOPT in order to match the GW170817 data, the hybrid EoS based on the soft SLy5 model favor larger $n_\fopt$ compared to the hybrid EoS based on the stiff PKDD model. The hybrid EoS based on the stiff PKDD model prefers the FOPT in the range of $n_\textrm{sat}\le n_{\fopt} \le 2n_\textrm{sat}$ while hybrid EoS based on the soft SLy5 model allows $1.5n_\textrm{sat}\le n_{\fopt} \le 3.5n_\textrm{sat}$ for the common domain of $\mu^*$.

In conclusion, for BNS, NSHS and BHS configurations, the largest correlations among the QM parameters are influenced by the stiffness of the nuclear EoS.
 
\section{Conclusions}
\label{Con}

In this study, we have explored nuclear and hybrid EoS, where the transition to QM is of first order and the QM EoS is linear. This general approach accurately approximates modern QM models. We have shown that the FOPT approach still depends 
on the stiffness nuclear EoS, influenced by the nuclear symmetry energy. 
Two representative models have been considered, the Skyrme SLy5 models for the soft nuclear EoS and the PKDD relativistic Lagrangian for the stiff nuclear EoS.

The uncertainties quantification for the FOPT parameters and the nuclear physics and astrophysics constraints are implemented within the Bayesian statistical approach. All EoS respect causality, stability, and the observational evidence that stable CS TOV masses are above $2M_\odot$. In this way, a total of about $400\ 000$ priors are considered.

We observe that the tension between models and data (NICER and GW) existing for nuclear models can be solved by considering hybrid EoSs built from the FOPT approach.

We found that nuclear stiffness impacts hybrid EoS and the nature of CS in GW170817: 
both soft and stiff nuclear EoS can produce hybrid EoS, suggesting that GW170817 is a BHS, while only soft nuclear EoS predicts GW170817 to be a BNS or NSHS system. In other words, if the nuclear EoS is stiff, GW170817 can only be a BHS. In addition, the $\tilde{\Lambda}$ measurement from GW170817 is best reproduced for HS built on hybrid stiff PKDD EoS.

Detailed analyses are performed in this study, and we show that the hybrid EoS based on stiff nuclear EoS requires softening, while the one based on soft nuclear EoS can be soft or stiff and still reproduce GW170817. We have compared globally our predictions to NICER measurements, but to simplify the discussion, we have not considered NICER measurements in our constraints.

Finally, we have shown that the correlation between $p_{\fopt}$ and $\mu^*_\QM$, as well as between $p_{\fopt}$ and $n_{\fopt}$ FOPT properties, is impacted by nuclear stiffness as strongly influenced by the nuclear symmetry energy.

We conclude on the important role played by nuclear stiffness on the properties of HS.
Since nuclear stiffness is strongly influenced by the nuclear symmetry energy, a precise determination of the symmetry energy, particularly its density dependence, is crucial for interpreting astrophysical observations and constraining the phase diagram of dense matter.  We plan, for instance, to explore more widely the symmetry energy by varying, e.g., the curvature of the symmetry energy $K_\sym$. Future work will benefit from incorporating a broader set of nuclear EoS and considering pulsar timing measurements (NICER, for instance) in the Bayesian constraints to further refine our understanding of the hadron-quark phase transition in neutron stars. 

\begin{acknowledgments}
This work is supported by the Scientific and Technological Research Council of Turkey (T\"{U}B\.{I}TAK) under project number MFAG-122F121 and the Yildiz Technical University under project number  FBA-2021-4229. J.M. and E.K. are supported by the CNRS-IN2P3 MAC masterproject and this work benefited from the support of the project RELANSE ANR-23-CE31-0027-01 of the French National Research Agency (ANR). J.M. benefits from the LABEX Lyon Institute of Origins (ANR-10-LABX-0066). 
\end{acknowledgments}

\bibliography{biblio}


\end{document}